\documentstyle[12pt]{article}
\begin{document}
\centerline{ SU(3) Clebsch-Gordan Coefficients\footnote{ published in
Romanian in St. Cerc. Fiz. {\bf 36}, 3 (1984)} }
\vskip1cm
\centerline{ Marius Grigorescu }
\begin{center}
Department of Applied Mathematics \\
The University of Western Ontario \\
London, ON, Canada N6A 5B7 \\
\end{center}
 \vskip1cm
Abstract: The purpose of this paper is to find out a set of general
recurrence formulas for the calculus of the $SU(3)$ Clebsch-Gordan
coefficients. The first six sections are introductory, presenting the
notations and general group theoretical methods applied to $SU(3)$. 
The following eight sections are devoted to a detailed treatment of the
carrier spaces of the irreducible representations and their direct 
product.\\[2cm]
{\bf I. Introduction} \\[.5cm] \indent
The first applications of the SU(3) group theory have occurred in nuclear
physics, following the attempts to describe the nuclear collective
properties in the frame of the shell model. The complex problem of the
states classification and determination of the energy spectrum was
simplified by using the dynamical symmetry of the
Hamiltonian. This is reflected by certain regularities observed in the
spectra, and indicates the existence of a set of operators which commute
with the many-body Hamiltonian and generate a Lie algebra. Because the
new "constants of motion" are not related to the already present
geometrical symmetries of the physical system, the spectrum shows an 
additional degeneracy. The subspace of the states with the same energy
carries a representation of this extended algebra, 
integrable to the Lie group of dynamical symmetry. The group $SU(3)$
represents the dynamical symmetry of the isotropic 3-dimensional harmonic
oscillator, whose degenerate energy eigenstates provide the basis of
a space of irreducible representation (irrep). The irreducible spaces and 
representations will be denoted by $V(P,Q)$ and $D(P,Q)$, respectively,
where $P$ and $Q$ are positive integers. The classification of the
harmonic oscillator states may be obtained by using the complete set of
commuting operators, whose eigenvalues are the indices of the basis
vectors of the $SU(3)$ irreducible representations \cite{1} ch. 13. The
classification of the states of the nucleon system in the harmonic
oscillator potential was obtained by Elliott \cite{2}, and it was applied
to the light nuclei. Bargmann and Moshinsky have considered in addition to
the harmonic oscillator potential also the residual two-particle
interaction \cite{3}- \cite{5}. This Hamiltonian is $SU(3)$-invariant, and
to obtain the eigenfunctions of a system consisting of 3 nucleons
it is necessary to know the $SU(3)$ Clebsch-Gordan
(CG) coefficients for the product $D(P,Q) \otimes D(P_1,0)$. \\ \indent
The coupling of the oscillator wave functions in the 2s-1d shell was
obtained
by Hecht \cite{6} by using the CG coefficients for the products 
\begin{equation}
D(P,Q) \otimes D(P_1,0)~~,~~ D(P,Q) \otimes D(P_1,1) ~~,~~ D(P,Q) \otimes
D(0,Q_1)~~. 
\end{equation}
For even-even nuclei was used with success the interacting boson
model \cite{7}. In this model only the symmetry of the valence nucleons is
considered, supposing that they are coupled in  $s$ and $d$- boson
pairs. At a fixed number of bosons the most general Hamiltonian of this
system has the $SU(6)$ symmetry, and the states are labeled by the
eigenvalues of the Casimir operators for a subgroup chain of
$SU(6)$. There are only three possible subgroup chains, one of them
starting with $SU(3)$. The energy
levels calculated by using this chain are in a good agreement with the
experiment for the nuclei with half-filled valence shell. In this case 
the $SU(3)$ symmetry  describes rotational spectra, and is independent of 
the degeneracy of the single-particle levels. \\ \indent
The importance of the simple Lie group theory for the elementary particle
physics was exposed in detail in ref. \cite{8}. The discovery of the
flavor quantum numbers such as the isospin, hypercharge, charm, conserved
by the strong interactions has lead to the introduction of the dynamical 
symmetry models based on $SU(2)$, $SU(3)$, and $SU(4)$, respectively.   
These symmetries are only approximate, because the particles
assigned to the multiplets have different masses. However, the
classification
is correct, because the symmetry breaking interactions do not change the
values of the internal quantum numbers. \\ \indent
In general, a hadronic model with $n_F$ constituent quarks (flavors) is
related to the $SU(n_F)$ symmetry. The $n_F$ quarks can be distinguished by the 
values of a set of $n_F$ flavor quantum numbers. If there are no charmed
particles, the hadronic systems are correctly described by the octet
model, based on the flavor symmetry group $SU(3)^F$. A basic reference to
this model and its applications is \cite{9}. In this reference several
tables of CG coefficients are included, and are used to calculate the
matrix elements of the mass
operator. This leads to a relationship between the strengths of
the coupling between the pseudoscalar meson octet and the barionic
currents in the case of the Yukawa coupling. The tables are also used to
express the hadronic wave functions in terms of quarks. \\ \indent
The $SU(3)$ invariance of the strong interactions leads also to useful 
equations for the calculus of the multiplet dependent
factor of the hadron scattering amplitude \cite{10}. In a many-quark
system, these amplitudes can be obtained only by knowing the CG series and 
coefficients for the direct product of two arbitrary representations. 
\\ \indent
In the octet model the barions are $L^\pi = 0^+$ states of a three quark
system, contradicting the expected behavior of a system of three
particles of spin $1/2$, obeying the Fermi-Dirac statistics. This puzzling
situation was solved by introducing the color quantum number taking three 
possible values, in addition to the flavor and Dirac spinor indices of the
relativistic quark wave function \cite{11}. The colored quarks are usual
spin $1/2$ Fermions, but unobservable, because by postulate, only the
"white" systems can be free. The system of three colored quarks is
antisymmetric to the permutation of the color indices, and is invariant to
their transformation by $SU(3)$. However, contrary to $SU(3)^F$,
the global color transformation group, $SU(3)^C$, is an exact
symmetry. Assuming that the quark Lagrangean is invariant not only to
global, but also to local $SU(3)^C$ transformations, it
is possible to formulate a Yang-Mills theory of the strong interactions,
the quantum chromodynamics \cite{12}. Although based on $SU(3)$, this 
theory does not rely on the $SU(3)$ irreducible representations, and the 
commutation relations between the gauge fields and currents are completely
specified by $su(3)$, the Lie algebra of $SU(3)$. \\ \indent
In certain physical situations the $SU(3)$ group is too restrictive, and
it is necessary to use larger semisimple Lie groups, as $SU(n)$. The
general theory of semisimple Lie groups was elaborated in the classical
works of Cartan and Weyl \cite{13}. The development of the quantum theory
of angular momentum has determined the intensive study of $SU(2)$. The
results obtained by Wigner \cite{14} and Racah \cite{15} in the theory of
the $SU(2)$ representation and irreducible tensor operators, have solved
practically all the problems of the atomic spectroscopy. The extension of
these results to $SU(n)$ was given in the series of papers \cite{16},
by Baird abd Biedenharn.  In ref. \cite{16}-II the method of boson
generators is applied to recover the general formulas of Gel'fand
and Zetlin  for the matrix elements of the generators. A particular 
attention is given to $SU(3)$, because in its study are encountered
features
characteristic for $SU(n)$ in general, but which are not revealed by the
simple structure of $SU(2)$. The last papers \cite{16}- IV,V concern the
calculus of the CG coefficients as matrix elements of the irreducible
tensor operators in the Gel'fand basis, by using the Wigner-Eckart
theorem. The main difficulty, represented by the occurrence of the
multiplicities, is solved by the classification of the tensor operators
using a conjugation operation. \\ \indent
The computational method used to obtain the $SU(3)$ and $SU(4)$ CG
coefficients was presented in ref. \cite{17}, \cite{18} and \cite{19},
respectively. An alternative
procedure, based on the Wigner-Eckart theorem, is presented in
ref. \cite{20}. There nonorthonormal isoscalar factors are calculated by  
using the recurrence formulas satisfyed by the matrix elements of the
irreducible tensor operators. \\ \indent
A general method for the calculus of the CG coefficients for every group
$SU(n)$, based on the properties of the projection operators associated to
the matrix elements of the irreducible representations is presented in 
ref. \cite{1}, ch. 7. The CG coefficients obtained this way are expressed
as linear combinations of integrals over the group parameters \cite{21}.
\\ \indent
The products of Eq. (1) have been reduced by Hecht, using the recurrence
formulas between the CG coefficients.  Isoscalar factors of
particular interest, as well as formulas for the general ones, have been 
derived  in \cite{22}, by using expansions of $SU(3)$ invariant
polynomials. \\ \indent
The results presented in these works solve completely the problems related 
to the $SU(3)$ irreducible representations and the related CG
coefficients. However, I have considered that a self-contained
presentation of some of these results, as well as of a simple and general
method to calculate the isoscalar factors, might be useful. The sections II-VI
are devoted to the basic definitions.  The notations used for the 
$SU(3)$ algebra, its fundamental and general tensor representations, as well as 
the CG series, are introduced. The sections VII-IX contain a detailed
study of the irrep spaces and their direct product, while in sect. X-XIV
the formulas obtained before are applied to
the explicit calculation of the CG coefficients in particular situations of 
interest. 
\\[.5cm]
{\bf II. The SU(3) group} \\[.5cm]
\indent The group $SU(3)$ consists of the linear unimodular
transformations
of the complex space $C^3$ which leave invariant the Hermitian bilinear
form 
\begin{equation}
<a,b> = a^*_1 b_1 + a^*_2 b_2 + a^*_3 b_3~~,~~a,b \in C^3~. 
\end{equation}
The matrices of the unitary linear operators associated to these
transformations in the basis
\begin{equation}
\{ x_j \in C^3, j=1,2,3 ; <x_j, x_k> = \delta_{jk} \}
\end{equation}
are unitary unimodular matrices $U$ which give a 3-dimensional
representation of the group $SU(3)$. An analytic structure on this group
can be introduced by considering the real and the imaginary part of the 9
complex elements of the matrix $U$ as analytical coordinates of a point
in the space $R^{18}$. These 18 real parameters are constrained by 10
algebraic relations determined by the conditions
\begin{equation}
det U=1 ~~,~~ UU^\dagger = I~~,
\end{equation}
such that only eight of them are independent. The remaining parameters
can be expressed as analytical functions of them, and therefore, $SU(3)$ 
is a real, eight-dimensional, analytical manifold.    \\ \indent   
 If $U_1,U_2 \in SU(3)$, then the coordinates of $U= U_1 U_2^{-1}$ are 
analytical functions of the coordinates of $U_1$ and $U_2$. Thus,
$SU(3)$ is a real Lie group. \\ \indent
By using the continuity of the matrix elements and Eq. (4) it can be shown
that $SU(3)$ is homeomorphic with a compact set in $R^8$ \cite{23}-IV. 
Moreover, it can be proved by recurrence that $SU(n)$ is simply connected
if $SU(n-1)$ is simply connected \cite{23}-(VIII sect. 4). As $SU(1) = \{
I \}$ is simply connected, then every $SU(n)$, and in particular $SU(3)$,
is simply connected.   \\ \indent
The coordinates on $SU(3)$,  $\{ q_j, j=1,8 \}$, are defined such
that the origin of the space $R^8$ corresponds to the unit matrix
$I$. Their choice is simplified
by the existence of the subgroups $SU(2)$ and $SUD(3) = D(3) \cap SU(3)$ (
consisting of the diagonal $3 \times 3$ unitary unimodular matrices), for 
which the coordinates are known. The four remaining parameters are chosen
usually in the form given by Nelson \cite{24}. Other parametrizations are
given in \cite{25}-\cite{27}. \\ \indent
Every $SU(3)$ matrix can be obtained from the identity $I$ by a series
of infinitesimal transformations corresponding to a continuous variation
of the group parameters, and can be represented in the form
\begin{equation}
U(q_1,...,q_8) = e^{ \frac{i}{2} \sum_{k=1}^8 q_k \lambda_k }~~.
\end{equation}
The parameters $q_k$, $k=1,8$ of this representation define a canonical
system of coordinates \cite{23}-(IX, sect. 3), and the matrices of the
generators of the infinitesimal transformations 
\begin{equation}
F_k = \frac{1}{2} \lambda_k~~,~~ k=1,8
\end{equation}
give a representation in $C^3$ of the real Lie algebra $su(3)$. These
matrices are the elements of a basis in the real linear space of the 
trace 0, $3 \times 3$ Hermitian matrices. Because $su(2) \subset su(3)$,
it is convenient to chose these matrices in the form given by Gell-Mann
\begin{equation}
\lambda_1=
\left[ \begin{array}{ccc}
0 & 1 & 0 \\
1 & 0 & 0 \\
0 & 0 & 0  
\end{array} \right]~,~
\lambda_2=
\left[ \begin{array}{ccc}
0 & -i & 0 \\
i & 0 & 0 \\
0 & 0 & 0  
\end{array} \right]~,~
\lambda_3=
\left[ \begin{array}{ccc}
1 & 0 & 0 \\
0 & -1 & 0 \\
0 & 0 & 0  
\end{array} \right]~,
\end{equation}
\begin{equation}
\lambda_4=
\left[ \begin{array}{ccc}
0 & 0 & 1 \\
0 & 0 & 0 \\
1 & 0 & 0  
\end{array} \right]~,~
\lambda_5=
\left[ \begin{array}{ccc}
0 & 0 & -i \\
0 & 0 & 0 \\
i & 0 & 0  
\end{array} \right]~,~
\lambda_6=
\left[ \begin{array}{ccc}
0 & 0 & 0 \\
0 & 0 & 1 \\
0 & 1 & 0  
\end{array} \right]~,
\end{equation}
\begin{equation}
\lambda_7=
\left[ \begin{array}{ccc}
0 & 0 & 0 \\
0 & 0 & -i \\
0 & i & 0  
\end{array} \right]~~,~~
\lambda_8= \frac{1}{ \sqrt{3}}
\left[ \begin{array}{ccc}
1 & 0 & 0 \\
0 & 1 & 0 \\
0 & 0 & -2  
\end{array} \right]~,~
\end{equation}
As there is an $SU(3)$ transformation which brings every $SU(3)$ matrix to
the form $SUD(3)$, the elements of $SU(3)$ belong to equivalence classes, 
such that every class contains only one element of
$SUD(3)$. Consequently, the characters for the elements of $SU(3)$ will be
analytical functions of the two variables which parameterize $SUD(3)$. An
invariant measure, normalized in the character space, can be found in 
\cite{28}. \\ \indent
The matrices $\lambda$ satisfy the relationship
%eq10
\begin{equation}
\lambda_i \lambda_j = \frac{2}{3} \delta_{ij} I + \sum_{k=1}^8 
(d_{ijk} + i f_{ijk}) \lambda_k~~.
\end{equation}
Here $d_{ijk}$ is a real, symmetric, tensor of order three,  $f_{ijk}$ is
also real, but antisymmetric, and $\delta_{ij}$ is the Kronecker
symbol. The non-vanishing components of these tensors are
\begin{equation}
f_{123}=1~~, f_{458} = f_{678} = \frac{ \sqrt{3}}{2}
\end{equation}
$$
f_{147}=f_{246}=f_{345}=f_{257}= - f_{156}= - f_{367}=
\frac{1}{2}~~, 
$$      
and
\begin{equation}
d_{118}= d_{228} = d_{338} = - d_{888} = \frac{1}{ \sqrt{3}}~~,~~
\end{equation}
$$
d_{448}= d_{558}=d_{668}=d_{778}= - \frac{1}{ 2 \sqrt{3}}~~,
$$
$$
d_{146} = d_{157} = d_{256} = d_{344}= d_{355} = - d_{247} = -
d_{366} = -d_{377} = \frac{1}{2}~~.
$$
By using Eq. (10) it is possible to derive the commutation and
anti-commutation relationships
\begin{equation}
[ F_i, F_j ] = i \sum_{k=1}^8 f_{ijk} F_k ~~,~~
\{ F_i, F_j \} = \frac{1}{3} \delta_{ij} I + \sum_{k=1}^8 d_{ijk} F_k~~.
\end{equation}
\indent The rank of the $su(3)$ algebra, defined as the dimension of the
maximal nilpotent Lie subalgebra, it is given by the dimension of the
$sud(3)$ subalgebra, which equals 2. The two commuting elements of $su(3)$
are $F_3$ and $F_8$, the basis in Eq. (3) being chosen such that their
matrices are diagonal. Their diagonal elements are related
to the quantum numbers characteristic to the systems with $SU(3)$
symmetry, and can be used to label the states. For the hadron 
classification the states are labeled by the eigenvalues $i_3$ and $y$ of
the operators $I_3$ and $Y$,
\begin{equation}
I_3 = F_3 ~~,~ Y= \frac{2}{ \sqrt{3}} F_8~~.
\end{equation}
The structure constants $c^i_{jk}$, $i,j,k=1,8$, defined by the 
commutation relations
\begin{equation}
[ i F_j, i F_k ] = \sum_{j=1}^8 c^l_{jk} (i F_l )
\end{equation}
are $c^l_{jk} = - f_{jkl}$. The Cartan metric tensor is defined by
\begin{equation}
g_{ij} = \sum_{k,l=1}^8 c^k_{il} c^l_{jk}~~,
\end{equation}
and has the simple expression $g_{ij} = - 3 \delta_{ij}$. Therefore the
Killing form
\begin{equation}
( X, Y) = \sum_{i,j=1}^8 g_{ij} X^i Y^j
\end{equation}
is non-degenerate, negative definite, and the algebra $su(3)$ is
semisimple. This means that $su(3)$ contains no commutative ideal, and
implies that $SU(3)$ is also a semisimple Lie group \cite{23}, Ch. XI,
sect. 4.  \\ \indent
The properties summarized above show that $SU(3)$ is a real,
compact, simply connected, semisimple Lie group. Such a group has no
proper connected invariant subgroup, but it may have a discrete one. For
$SU(3)$ this subgroup is
\begin{equation}
Z_3 = \{ I, e^{ \frac{2}{3} i \pi } I, e^{ \frac{4}{3} i \pi } I \}~~.
\end{equation}  
The factor group $SU(3) / Z_3$ is obtained by identifying the elements of
$SU(3)$ which are different by a factor           
$e^{ \frac{2}{3} i \pi }$ or $e^{ \frac{4}{3} i \pi }$, and is triply
connected. This factor group is important in the elementary particle
physics, because only the hadron states assigned to its irreps have
integer charge and hypercharge.   \\[.5cm]
{\bf III. The su(3) algebra } \\[.5cm] \indent
The linear representations theory of the semisimple Lie algebras
$L$, and the corresponding groups, rely on the decomposition of $L$ in a
direct sum of invariant subspaces with respect to the restriction of the
adjoint representation
\begin{equation}
X \rightarrow ad_X = [X, * ]~~,~~ X \in L
\end{equation}
to the nilpotent subalgebra
\begin{equation}
H = \{ X, Y \in L / (ad_X)^k Y =0, 0<k \in Z \}~~.
\end{equation}
An important example is provided by the Gauss decomposition, $L = N^+ + H
+ N^- $, where $N^{\pm}$ are nilpotent subalgebras, while $N^{ \pm} + H$
are solvable \cite{1}, ch. 1, sect. 6. \\ \indent
The $su(3)$ algebra and the $SU(3)$ group have no Gauss decomposition
because are semisimple compact ( \cite{1}, ch. 3, sect. 6),  but they
represent compact real forms of the algebra $sl(3,C)$, respectively the
group $SL(3,C)$, which are semisimple complex, and have a Gauss
decomposition. The representations of $su(3)$ and $SU(3)$ which appear in
applications are completely determined by the representations of
$sl(3,C)$, respectively $SL(3,C)$ by the unitary Weyl trick \cite{23}, XI,
XII. \\ \indent
The Lie algebra $sl(3,C)$, ($A_2$), is the subalgebra of $gl(3,C)$
generated by the traceless $3 \times 3$ matrices. A basis of $sl(3,C)$ is
represented by the matrices $F_k$, $k=1,8$ introduced in the previous
section, but this choice is not appropriate  for the Gauss
decomposition. Instead, it is convenient to express the basis elements  
as linear combinations of the basis of the $gl(3,C)$ algebra. The Weyl
basis of $gl(3,C)$  is represented by nine $3 \times 3$ matrices 
$\{ e_{ik}, i,k=1,3 \}$ which have a single non-vanishing element,
equal to 1,
\begin{equation}
(e_{ik})_{\alpha \beta} = \delta_{i \alpha} \delta_{k \beta}~~.
\end{equation}
Their commutator
\begin{equation}
[ e_{ik}, e_{jl} ] = \delta_{kj} e_{il} - \delta_{il} e_{jk}~~,   
\end{equation}
is a matrix of $sl(3,C)$, and therefore the same commutation relations 
will be satisfied by the nine traceless matrices $A^i_k = e_{ik} -
\delta_{ik} I /3$,
\begin{equation}
[ A^i_k, A^j_l ] = \delta_{kj} A^i_l - \delta_{il} A^j_k ~~.  
\end{equation}
\indent For applications in nuclear physics it is useful to remark that
operators $e_{ik}$ which satisfy the commutation relations (22)    
can be constructed by using boson operators. Thus, if 
$a^\dagger_i$ and $a_i$, $i=1,2,3$, denote the creation and annihilation
boson operators, ($[a_i, a_k^\dagger] = \delta_{ik}$, $
[a_i, a_k]= [a_i^\dagger, a_k^\dagger] =0$),  which appear in
the Hamiltonian of the isotropic harmonic oscillator,
\begin{equation}
h_0 = \hbar \omega \sum_{i=1}^3 (a^\dagger_i a_i + \frac{1}{2} )~~,
\end{equation}
then $e_{ik} = a^\dagger_i a_k$ satisfy Eq. (22), and $[e_{ik}, h_0] =0$.
\\ \indent
The relationship between the two basis sets for $sl(3,C)$, 
$\{ A^i_k ; i,k=1,3$ and $\{F_k, k=1,8 \} $ takes a simple form in terms 
of the complex combinations  
\begin{equation}
I_{\pm} = F_1 \pm i F_2 ~~,~~ K_{\pm} = F_4 \pm i F_5 ~~,~~ 
L_{\pm} = F_6 \pm iF_7~.
\end{equation}
By using these new operators, the relationship between the two basis sets
is given by the equations
\begin{equation}
A^1_1 = \frac{1}{ \sqrt{3}} F_8 + F_3 ~~,~~ A^2_2 = \frac{1}{ \sqrt{3}}
F_8 - F_3 ~~,~~ A^3_3 = - \frac{2}{ \sqrt{3}} F_8~,
\end{equation}
$$
A^1_2 = I_+~~,~~A^1_3 = K_+~~,~~A^2_3= L_+~~,
$$
\begin{equation}
A^2_1 = I_-~~,~~ A^3_1 = K_-~~,~~A^3_2= L_-~~.  
\end{equation}
\indent The algebra $sl(3,C)$ decomposes in the Cartan subalgebra 
$H$, which is generated by the elements $F_3$ and $F_8$, and the 
subalgebras $N^\pm$, generated by $I_\pm$, $K_\pm$ and $L_\pm$.
By using the expansion
\begin{equation}
F_\rho = \sum_{k=1}^3 \Phi_k ( \rho) e_{kk}
\end{equation}
where $\rho$ is 3 or 8, the $sl(3,C)$ commutation relations in the
Cartan-Weyl basis are expressed by
\begin{equation}
[ F_\rho , A^j_k ] = \alpha _{jk} ( \rho ) A^j_k
\end{equation}
\begin{equation}
[ A^i_k,A^j_l] = \delta_{kj} A^i_l - \delta_{il} A^j_k~~,i \ne k~,~j \ne
l~~.
\end{equation}
\indent The coefficients $\alpha_{jk}$ are linear functions on $H$, 
defined by
\begin{equation}
\alpha_{jk} ( \rho ) = \Phi_j ( \rho ) - \Phi_k ( \rho)~~, 
\end{equation}
and the set
\begin{equation}
\Delta = \{ \alpha_{jk} ; j,k=1,2,3 \}
\end{equation}
represents the root system. \\
\indent The linear combinations of the roots generate the dual space of
$H$, denoted $\tilde{H}$. If $\alpha_{12}$, $\alpha_{13}$ and
$\alpha_{23}$ are chosen positive, then $\alpha_{12}$ and 
$\alpha_{23}$ are the simple roots. The new structure constants
\begin{equation}
C^{(ik)}_{ \rho (jl)} = \alpha_{jl} ( \rho ) \delta_{ij} \delta_{kl}
\end{equation}
give the restriction of the metric tensor to the Cartan subalgebra   
\begin{equation}
g_{ \rho \sigma} = (F_\rho , F_\sigma ) = \sum_{j,l=1}^3 
\alpha_{jl} ( \rho) \alpha_{jl} ( \sigma) = 3 \delta_{ \rho \sigma}~~,
\end{equation}
with $\rho, \sigma=$ 3 or 8. Thus, the restriction of the Killing form 
to $H$ is non-degenerate, and for every  $\alpha \in \tilde{H}$
exists an unique element $h_\alpha \in H$,  denoted in the following also
by $\alpha$, such that $\forall h \in H$ 
\begin{equation}
\alpha (h) = (h_\alpha , h)~~,~~( \alpha, \beta ) \equiv ( h_\alpha,
h_\beta )~~, \alpha, \beta \in \tilde{H}~~.
\end{equation}
Reciprocally, this relationship associates linear functions to the
elements of $H$. The linear functions $\alpha_3$ and $\alpha_8$ associated
by Eq. (29) to $F_3$ and $F_8$ are identically zero. However, Eq. (34) and
(35) lead to a covariant orthogonal basis $\hat{g}_\sigma$, $\sigma=$ 3 or
8, in $\tilde{H}$, defined by $\hat{g}_\sigma ( \rho) = g_{ \sigma \rho}
$. The covariant coordinates of the roots from $\Delta$ in this basis 
are given by Eq. (35). \\ \indent
Another covariant basis in $\tilde{H}$ is represented by the simple roots.
These have the length  $1/ \sqrt{3}$ with respect to the Killing form,
and span an angle of 120 degrees. The matrices associated by Eq. (35) to
these roots are
\begin{equation}
\alpha_{12} = \sum_{ \rho, \sigma} g^{\rho \sigma} \alpha_{12} ( \rho)
F_\sigma = \frac{1}{3} F_3 = \frac{1}{6} 
\left[ \begin{array}{ccc}
1 & 0 & 0 \\
0 & -1 & 0 \\
0 & 0 & 0  
\end{array} \right]~~,
\end{equation}
\begin{equation}
\alpha_{23} = \sum_{ \rho, \sigma} g^{\rho \sigma} \alpha_{23} ( \rho)
F_\sigma =  \frac{1}{6} (\sqrt{3} F_8 - F_3)  = \frac{1}{6} 
\left[ \begin{array}{ccc}
0 & 0 & 0 \\
0 & 1 & 0 \\
0 & 0 & -1  
\end{array} \right]~.
\end{equation}
\indent The contravariant vectors $\alpha^\mu$, $\mu=$ 12 or 23 associated
to the simple roots are defined by the relationship
\begin{equation}
\frac{ ( \alpha^\mu , \alpha_\nu )}{( \alpha_\nu, \alpha_\nu)} =
\frac{1}{2} \delta_{ \mu \nu}~~,
\end{equation}
and the coordinates of an arbitrary element $M \in \tilde{H}$ in the
contravariant basis are 
\begin{equation}
m_\mu = 2 \frac{ (M, \alpha_\mu)}{(\alpha_\mu, \alpha_\mu)}~~.
\end{equation}
\\[.5cm]
{\bf IV. Fundamental representations } \\[.5cm] \indent
The irreducible representations of a Lie group $G$ which admits a
Gauss decomposition of the form $G = Z^- D Z^+$, with $Z^\pm$, $D$,
generated by $N^\pm$, $H$, respectively, are induced by the 
one-dimensional representations (characters) $\tau ( \delta)$, $\delta \in
D$ of the subgroup $D$. Let $R_g$, $g \in G$, a representation of $G$ on
the finite dimensional linear space $V$. The elements of $V$
which are eigenstates of the operators $R_\delta$, $\delta \in D$ are
called weight vectors. In particular, the weight vectors $\vert M>$,
\begin{equation}
R_\delta \vert M> = \tau_M ( \delta) \vert M>~~,~~ \forall \delta
\in D
\end{equation}
which remain invariant to the action of the subgroup $ Z^+$,
\begin{equation}
R_z \vert M > = \vert M >~~, \forall z \in Z^+~~,
\end{equation}
are called highest weight vectors. \\ \indent
In the linear irreps theory of the group $GL(n,C)$ are proved the
following important theorems: \\
{\bf Th. I.} Every carrier space $V$ of a finite dimensional irrep $R$ of
$GL(n,C)$ contains an unique highest weight vector, cyclical for $V$. \\
{\bf Th. II.} The irrep  induced by the character $\tau_M$ occurrs in the 
decomposition of an reducible representation on the space
$V$ with a multiplicity equal to the number of the highest weight vectors
$\vert M>$ contained by $V$ \cite{23} ch. VI, sect. 3.1. \\
{\bf Th. III.} The analytical complex inductive characters for the
representations of the group $GL(3,C)$ have the expression
\begin{equation}
\tau ( \delta ) = \gamma_1^{m_1} \gamma_2^{m_2} \gamma_3^{m_3} ~~,
\delta = 
\left[ \begin{array}{ccc}
\gamma_1 & 0 & 0 \\
0 & \gamma_2 & 0 \\
0 & 0 & \gamma_3  
\end{array} \right] \in D ~~,
\end{equation}
with $m_1 \ge m_2 \ge m_3$, integers \cite{1} ch. 8 sect. 3. \\   
\indent The finite dimensional irreps of the group
$SU(3)$ can be realized on the irrep spaces of its complex extension,
$SL(3,C)$. The restriction of the characters $\tau$ of $D \subset
GL(3,C)$ to $SUD(3) \subset D$ defines a character $\tau^0$ which
specifies completely the irreps of $SU(3)$. This restriction is obtained
by considering only matrices $\delta$ with elements $\gamma_i$, $i=1,2,3$
such that
\begin{equation}
\vert \gamma_i \vert =1 ~~,~~\gamma_1 \gamma_2 \gamma_3 =1~~.
\end{equation}
\indent If the simple roots are $\alpha_{12}$ and $\alpha_{23}$, then every
matrix $d \in SUD(3)$ can be written as 
\begin{equation}
d = e^{i h}
~~,~~
h=t^{12} \alpha_{12} + t^{23} \alpha_{23}~~, 
\end{equation}
where the real parameters $t^{12}$ and $t^{23}$ are  
\begin{equation}
t^{12} = 2 \frac{ (h, \alpha^{12})}{(\alpha_{12}, \alpha_{12})}~~,~~
t^{23} = 2 \frac{ ( h, \alpha^{23})}{( \alpha_{23}, \alpha_{23})}~~.
\end{equation}
Replacing $\alpha_{12}$ and $\alpha_{23}$ from Eq. (44) by the expressions
of Eq. (36), (37), the matrix elements of $d$ take the form
\begin{equation}
\gamma_1 = e^{ \frac{i}{6} t^{12} } ~~,~~ \gamma_2 = e^{ - \frac{i}{6} 
( t^{12} - t^{23})} ~~,~~ \gamma_3 = e^{ - \frac{i}{6} t^{23}}~~.
\end{equation}
These elements satisfy the conditions of Eq. (43), and therefore,
by {\bf Th. III}, the inductive character can be written in the form
\begin{equation}
\tau^0_{\underline{M}}  ( d) = e^{i ( \underline{M}, h)  }~~,
\end{equation} 
with $\underline{M} \in \tilde{H}$ defined by $\underline{M} =
(m_1-m_2) \alpha^{12} + (m_2 - m_3) \alpha^{23} $. The element
$\underline{M}$ of $\tilde{H}$ is called highest weight. Therefore,
every irrep of $SU(3)$ is completely specified by two non-negative
integers, $P= m_1-m_2$ and $Q= m_2 -m_3$, and is denoted $D(P,Q)$. 
The carrier space of this irrep will be denoted in the following by
$V(P,Q)$. By using the same notation for the elements of the $su(3)$ 
algebra and the corresponding representation operators in $V(P,Q)$,   
the expansion of Eq. (40) and (41) for $SU(3)$ near the identity
leads to the equations
\begin{equation}
\alpha_{12} \vert \underline{M}> = ( \alpha_{12}, \alpha^{12} ) P \vert 
\underline{M} >~~,~~ 
\alpha_{23} \vert \underline{M}> = ( \alpha_{23}, \alpha^{23} ) Q \vert 
\underline{M} >~~,
\end{equation}
and, respectively
\begin{equation}
I_+ \vert \underline{M} > =0 ~~,~~ L_+ \vert \underline{M} > =0 ~~.
\end{equation}
Thus, according to Eq. (36), (37), $\vert \underline{M}>$ is an 
eigenvector of $F_3$ and $F_8$, and by Eq.  (14), of $I_3$ and $Y$, with
the eigenvalues
\begin{equation}
(i_3)_{\underline{M} }= \frac{P}{2} ~~,~~ ( y)_{\underline{M}} = 
\frac{ P  + 2 Q}{3} ~~.
\end{equation}
Sometimes it is convenient to choose $\alpha_{13}$ and $\alpha_{32}$
as simple roots. In this case, the highest weight vector, denoted by
$\vert M>$,  will satisfy the equations
\begin{equation}
K_+ \vert M> =0 ~~,~~ L_- \vert M> =0~~,
\end{equation}
and the highest weight $M = P \alpha^{13} + Q \alpha^{32}$ is the
reflection of the weight $\underline{M}$ with respect to the contravariant
vector $\alpha^{13} = \alpha^{12}$. The quantum numbers $i_3$ and
$y$ labeling $\vert M>$ are 
\begin{equation}
(i_3)_M = \frac{P+Q}{2} ~~,~~ ( y)_M = \frac{ P  - Q}{3} ~~.
\end{equation}
\indent A basis in $V(P,Q)$ is represented by the eigenvectors $\vert m>$
of the operators $I_3$ and $Y$. The couple $(i_3,y)$ of the eigenvalues
labels the vectors $\vert m>$, and will be denoted in the first part of
this work by $m$. These eigenvalues are related to the  
components $m_{13}$ and $m_{32}$ of the weight $m$ in the contravariant
basis by
\begin{equation}
i_3 = \frac{ m_{13} + m_{32} }{2}~~,~~ y= \frac{ m_{13} -m_{32} }{3}~~.
\end{equation}
According to {\bf Th. I}, all basis vectors $\vert m> \equiv \vert i_3,y>$
can be obtained by the application of the operators $L_+$, $I_-$ and
$K_-$ on the highest weight vector $\vert (P+Q)/2, (P-Q)/3>$. \\ \indent
The numbers $P$ and $Q$ are related to the eigenvalues $f$ and $g$ of
the two Casimir operators $F$ and $G$ \cite{1} ch. 9 sect. 4 defined by
\begin{equation}
F= \sum_{k=1}^8 F_k^2 = \frac{1}{2} \sum_{i,k=1}^3 A^i_k A^k_i ~~,~~
G= \frac{1}{2} \sum_{i,k,l=1}^3 (A^i_l A^k_i A^l_k + A^l_i A^i_k A^k_l)~,
\end{equation}
such that
\begin{equation}
f= \frac{ P^2 + PQ + Q^2}{3} + P+Q
\end{equation}
$$
g= \frac{1}{9} (P-Q)(2P+Q+3)(P+2Q+3)~~.
$$
\indent The representations $D(1,0)$ and $D(0,1)$ are 3-dimensional,
non-equivalent, and are called fundamental representations. These have
the highest weights $\alpha^{13}$ and $\alpha^{32}$, respectively, and
every representation can be constructed  by the decomposition of their
multiple direct product. \\ \indent
By applying the commutator $[F_\rho, A^j_k ]$, $\rho=3,8$, of Eq. (29) to
the weight vector $\vert m>$, and then using the eigenvalues equations
\begin{equation}
F_\rho \vert m> = m_\rho \vert m > ~~,~~ m_3 = i_3~,~ m_8 = \frac{ 
\sqrt{3}}{2} y~~,
\end{equation}
one can find the relation
\begin{equation}   
F_\rho A^j_k \vert m> = (m_\rho + \alpha_{jk} ( \rho) ) A^j_k \vert m>~~,
\end{equation}
which gives the weight of the vector $A^j_k \vert m>$. \\ \indent
The representation matrices of the $su(3)$ generators in the space
$V(1,0)$  have the form of Eqs. (7),(8),(9), and the basis consists of the
highest weight vector $x_1 = \vert \frac{1}{2}, \frac{1}{3} >$ and 
\begin{equation}
x_2 = I_- x_1 = \vert - \frac{1}{2}, \frac{1}{3} > ~~,~~
x_3 = K_- x_1 = \vert 0, - \frac{2}{3} > ~~. 
\end{equation}
The weights $i_3,y$ of the vectors $x_1$, $x_2$, $x_3$, can be represented
on the root diagram by the vertices 1,2,3, respectively, of the triangle 
shown in Fig. 1. \\ \indent
The representation matrices of the $SU(3)$ group on $V(1,0)$ are just the
elements of the group of Eq. (5). These act by transforming the vectors
$x_j$, $j=1,2,3$ in $\underline{x}_j$, 
\begin{equation}
\underline{x}_j = \sum_{k=1}^3 U^k_j x_k ~~.
\end{equation}
\indent A new three-dimensional irreducible representation is obtained by
the complex conjugation of Eq.  (59), 
\begin{equation}
\underline{y}^j = \sum_{k=1}^3 (U^k_j)^* y^k ~~.
\end{equation}
Here $y^j=x_j^*$ are contravariant basis vectors which generate the 
representation space $V(1,0)^*$. The matrices of the $SU(3)$ elements
in the $D(1,0)*$ representation are 
\begin{equation}
U \vert_{(1,0)^*} = ( U \vert_{(1,0)} )^* = e^{ - \frac{i}{2} \sum_{k=1}^8 
\lambda_k^* a_k}~~,
\end{equation}
which shows that the matrices of the $su(3)$ generators are
\begin{equation}
 F_i \vert_{(1,0)^*} = - ( F_i \vert_{(1,0)} )^*~~.
\end{equation}  
The matrices of the operators $F_\rho \vert_{(1,0)^*}$ and $A^i_k \vert_{
(1,0)^*} $ can be found by using Eq. (25), such that
\begin{equation}
F_\rho \vert_{(1,0)^*}= - (F_\rho \vert_{(1,0)})^*~~,~~A^i_k \vert_{
(1,0)^*}= - (A^k_i \vert_{(1,0)})^* ~~.
\end{equation}
Here the matrices $F \vert_{(1,0)}$ and $A^i_k \vert_{(1,0)}$ are real,
and therefore 
\begin{equation}
F_\rho \vert_{(1,0)^*}= - F_\rho \vert_{(1,0)}~~,~~A^i_k \vert_{
(1,0)^*}= - A^k_i \vert_{(1,0)} ~~.
\end{equation}
The weight diagram of the vectors $y^k$ is obtained by reflection with
respect to the origin of the weight diagram for $x_k$, $k=1,2,3$, and
coincides with that of the basis in $V(0,1)$. If the basis vectors are
labeled by their weights, such that $x_m$ denotes the basis
vectors of Eq. (58), and $y^m$ the vectors $y^k$, then
\begin{equation}
y^{-m} = (x_m)^*~~.
\end{equation}
By convention, the matrix elements of the operators $I_{\pm}$ and
$K_{\pm}$ in the canonical basis of  $V(P, Q)$ are chosen to be positive
\cite{9}. This convention is violated by the basis $\{ y^i, i=1,2,3 \}$, 
because, as it follows from Eq. (64), 
\begin{equation}
I_+ y^1= - y^2~,~I_- y^2 = - y^1~,~K_+y^1 = -y^3~,~ K_-y^3=-y^1~.
\end{equation}
However, a canonical basis in $V(0,1)$, denoted  $\{ \eta_i, i=1,2,3 \}$
can be obtained from $\{ y^i, i=1,2,3 \}$ by the transformation
\begin{equation}
\eta_1 = - y^1~~,~~ \eta_2 = y^2 ~~,~~ \eta_3 = y^3 ~~,
\end{equation}
or
\begin{equation}
\eta_{m'} = \sum_{m} G^m_{m'} x_m^* = \sum_{m} G^m_{m'} y^{-m}~~.
\end{equation}
The transformation matrix $G$, as well as the basis vectors, is defined   
up to a phase factor. The choice of Eq. (67) corresponds to 
\begin{equation}
G^m_{m'} = (-1)^{ \frac{1}{3} + e_m} \delta_{m', -m}~~,
\end{equation}
where $e_m= (i_3 + y/2)_m$ is the electric charge of the state $m=(i_3,y)$
\cite{29}. \\ \indent
In the following, the spaces $V(1,0)$ and $V(0,1)$ will be denoted also by
$V(3)$ and $V(3^*)$, respectively, because they are 3-dimensional. 
\\[.5cm]
{\bf V. Tensor representations } \\[.5cm] \indent
{\bf Def. I.} The object $T_{i_1,...,i_P}$ is called covariant  tensor of
rank
$P$ with respect to $SU(3)$ if at the action of $U \in SU(3)$ 
has the transformation law
\begin{equation}
\underline{T}_{i_1,...,i_P} = U^{k_1}_{i_1} ... U^{k_P}_{i_P}
T_{k_1,...,k_P}~~.
\end{equation}
Here and in the following the summation convention of the repeated indices 
is considered. \\ \indent
{\bf Def. II.} The object $T^{i_1,...,i_Q}$ is called contravariant
tensor of rank $Q$ with respect to $SU(3)$ if at the action of $U \in
SU(3)$ has the transformation law
\begin{equation}
\underline{T}^{i_1,...,i_Q} = (U^{k_1}_{i_1})^* ... (U^{k_Q}_{i_Q})^*
T^{k_1,...,k_Q}~~.
\end{equation}
The representations determined by such transformation formulas
are called tensor representations, and are presented in detail in 
\cite{1} ch. 10, sect. 2. The components of the tensors {\bf I} and {\bf 
II} can be considered as elements of a $3^P$, respectively $3^Q$ -
dimensional space.
Mixed tensors can be obtained by the direct product of the covariant and
contravariant tensors. These spaces carry $SU(3)$ representations which are
in general reducible. Therefore, of a particular interest are those
tensors, called irreducible, whose components are the basis elements
in spaces of $SU(3)$ irreducible representation. \\ \indent
The action of $SU(3)$ defined by Eq.  (70),(71) commutes with the action of the
permutation group on the set of tensor indices. Therefore, irreducible
are only the tensors which, as functions of indices, provide an irreducible
representation of the permutation group. Such tensors are obtained by
linear combinations of the type - {\bf I} and {\bf II} tensors. The
resulting tensor should be either symmetric, or antisymmetric with respect
to the permutation of well- defined subsets of indices. \\ \indent Objects
with the transformation properties of the tensors {\bf I} and {\bf II} are
represented by the set of basis elements in the spaces obtained by the
multiple direct product of the spaces $V(3)$ and $V(3^*)$,
respectively. The spaces  
\begin{equation}
V(P) = V^1(3) \otimes ... \otimes V^P (3)~~,~~
V(Q) = V^1(3^*) \otimes ... \otimes V^P (3^*)~~,
\end{equation}
have as basis elements
\begin{equation}
(a) ~ T_{i_1 ... i_P} = x^{(1)}_{i_1} ... x^{(P)}_{i_P}~,
(b) ~ T^{j_1 ... j_Q} = y^{(1) j_1} ... x^{(Q) j_Q}~.
\end{equation}
The generators of the infinitesimal transformations in $V(P)$ have the form
\begin{equation}
F_k = F^{(1)}_k \otimes I^{(2)} \otimes ... \otimes I^{(P)} + ...
I^{(1)} \otimes I^{(2)} \otimes ... \otimes F^{(P)}_k~,
\end{equation}
where $F^{(i)}_k$, $k=1,8$,  and $I^{(i)}$ denote the $su(3)$ generator,
and, respectively, the unit operator, in the space $V^i(3)$. Similarly
is defined the action on $V(P)$ of the $su(3)$ operators $\{ A^i_k,
i,k=1,2,3 \}$. \\ \indent
The direct product of the highest weight vectors $x^{(k)}_1$, $k=1,...,P$
give the element of the tensor $T^{ \otimes_P} $
\begin{equation}
T^{ \otimes_P}_{11...1} = \vert \underline{M} > = x^{(1)}_1
... x^{(P)}_1~~,
\end{equation}
which is a highest weight vector in $V(P)$, with 
$(i_3,y)_{ \underline{M} }= (  P/2 , P/3 )$. The expressions of the
representation operators $I_-$, $K_-$, are similar to $F_k$ of Eq. (74),
and are symmetric with respect to $I^{(k)}_-$, $K^{(k)}_-$,
$k=1,P$. Therefore, their action on the highest weight component
$T^{\otimes_P}_{11...1}$ generates the components
of a symmetric, irreducible, covariant tensor. The symmetry with respect
to the permutation of the lower indices allows to replace the Kronecker
product of Eq. (73) by the Young product.  In the Young product, the 
vectors $x^{(k)}_i$ and $x^{(j)}_i$, from $V^k(3)$ and $V^j(3)$, with $k
\ne j $, are supposed to be the same, and are both denoted by $x_i$. This 
procedure leads to a mapping $T^{ \otimes_P} \rightarrow T^P$, where $T^P$
is a polynomial of degree P in three variables having the
same transformation properties under $SU(3)$ as the basis elements of
$V(3)$,
\begin{equation}
T^P_{i_1...i_P} = x_{i_1} ... x_{i_P} = (x_1)^{p_1} (x_2)^{p_2} (x_3)^{p_3}~~,
\end{equation}
where $p_1$, $p_2$, $p_3$ are non-negative integers such that
\begin{equation}
p_1+p_2+p_3=P~~.
\end{equation}
The components of  $T^P$ are eigenvectors of the operators
$F_3$ and $F_8$, and generate an orthogonal basis in $V(P,0)$. Similarly
it is possible to obtain irreducible representations equivalent to
$D(0,Q)$, on spaces generated by contravariant symmetric tensors,
\begin{equation}
T^{ Q j_1 ... j_Q} = y^{j_1} ... y^{j_Q} = (y^1)^{q_1} (y^2)^{q_2}
(y^3)^{q^3}~~,
\end{equation}
 where $q_1$, $q_2$, $q_3$ are non-negative integers such that
\begin{equation}
q_1+q_2+q_3=Q~~.
\end{equation}
 The $su(3)$ generators act on the tensors $T^P$ and $T^Q$ as differential
operators, completely defined by their action in the spaces $V(3)$
and $V(3^*)$.  \\ \indent
The dimension of the space $V(P,0)$ equals the number of the components of
the tensor $T^P$. This is given by the number of partitions of $P$ in
$p_1$, $p_2$, $p_3$, such that
\begin{equation}
\dim V(P,0) = N_P = \sum_{p_1=0}^P \sum_{p_2=0}^{P-p_1} 1 = 
\frac{(P+1) (P+2)}{2}~~.
\end{equation}
Similarly can be obtained the dimension of the space generated by the
tensor $T^Q$, 
\begin{equation}
\dim V(Q,0)^* = \dim V(0,Q)=N_Q = \frac{(Q+1) (Q+2)}{2}~~.
\end{equation}
\indent Representations which are equivalent to $D(P,Q)$ can be obtained 
by the decomposition of the direct product $D(P,0) \otimes D(Q,0)^*$. The
basis of the product space has the form 
\begin{equation}
T^{j_1...j_Q}_{i_1...i_P}= (x_1)^{p_1}(x_2)^{p_2}(x_3)^{p_3}(y^1)^{q_1}
(y^2)^{q_2}(y^3)^{q_3}~~,
\end{equation}
with
\begin{equation}
\sum_{i=1}^3p_i=P~~,~~\sum_{i=1}^3q_i =Q~~.
\end{equation}
This tensor is symmetric with respect to the permutation of the upper, as
well as of lower indices. By contraction with the invariant tensor
$\delta^i_j \equiv \delta_{ij}$, one obtains the sequence of tensors
$T^{(k)}$,
\begin{equation}
T^{(k) j_{k+1}...j_Q}_{i_{k+1}...i_P} = \delta^{i_k}_{j_k}
T^{(k-1) j_{k}...j_Q}_{i_{k}...i_P}~~.
\end{equation}
The spaces $V^{(k)}$ generated by them are all $SU(3)$ invariant, and
satisfy the relationship
\begin{equation}
V(P,0) \otimes V(Q,0)^* = V^{(0)} \supset V^{(1)} \supset ... \supset
V^{(n)}~~,~~n= \min(P,Q)~.
\end{equation}
The orthogonal complement of the space $V^{(k+1)}$ in $V^{(k)}$ with
respect to the scalar product defined by the contraction of the indices is
represented by the linear combinations of the tensors from $V^{(k)}$ which
are traceless with respect to any contraction, and is denoted by
$V^{(k)}_0$. Thus,
\begin{equation}
V^{(k)}= V^{(k)}_0 \oplus V^{(k+1)}~~,
\end{equation}
and
\begin{equation}
V(P,0) \otimes V(Q,0)^* = \sum_{k=0}^n V_0^{(k)} = 
\sum_{k=0}^n V(P-k,Q-k)~~.
\end{equation}
The representation on the space $V^{(0)}_0$ is equivalent to $D(P,Q)$,
and
\begin{equation}
\dim V(P,Q) = \dim V^{(0)} - \dim V^{(1)} = N_PN_Q- N_{P-1}N_{Q-1}=
\end{equation}
$$
= \frac{ (P+1) (Q+1) (P+Q+2)}{2} ~~.
$$
This is a special case of the general Weil's formula for
the dimension of the irreps for every simple Lie group \cite{23} ch. X
sect. 13.4, \cite{1} ch. 8 sect. 8. \\ \indent
The weight diagram of $V(P,Q)$ is obtained by substracting from the
highest weight $(i_3,y)_M= ( (P+Q)/2, (P-Q)/3 )$ linear combinations
with integer coefficients of the simple roots $\alpha_{12}$,
$\alpha_{13}$, 
and $\alpha_{32}$. This diagram has three symmetry axes, and is
bounded by the polygonal line drawn in Fig. 2. The number of weights of
this diagram is smaller than $\dim V(P,Q)$, and to label the states it is
necessary to find additional operators, which commute with $I_3$ and $Y$,
but not with all $A^i_k$, $ i \ne k$. \\ \indent
The problem of labeling the states of the $SU(n)$ irrep spaces is solved 
by the canonical factorization $SU(n) \supset U(1) \otimes SU(n-1)$
\cite{16}-II. The additional operators are in this case the Casimir
operators of the subgroups $SU(k)$, $k=2,...,n-1$. For $SU(3)$ the
additional operator can be chosen as the Casimir operator $I^2$ of the
$su(2)$ subalgebra $\{ I_-,I_3, I_+ \}$, 
\begin{equation}
I^2 = F_1^2+F_2^2+F_3^2~~,
\end{equation}
which satisfies 
\begin{equation}
[I^2,F_3]=0~~,~~[I^2,Y]=0~~,~~[I^2,I_{\pm}]=0~~.
\end{equation}       
\indent The basis vectors of the space $V(P,Q)$ can be labeled by the
eigenvalues of the operators which form a complete set, in this case
$F$, $G$, $I^2$, $I_3$ and $Y$. However, in applications, instead of the
eigenvalues $f$ and $g$ is more convenient to use the integers $P$ and
$Q$. The eigenvalue equations for the labeling operators have the known
form 
\begin{equation}
I^2 \vert PQii_3y> = i(i+1) \vert PQii_3 y>~~,
\end{equation}
\begin{equation}
I_3 \vert PQii_3y> = i_3 \vert PQii_3 y>~~,
\end{equation}
\begin{equation}
Y \vert PQii_3y> = y \vert PQii_3 y>~~,
\end{equation}
and the vectors of the canonical basis are subject to the  normalization
condition 
\begin{equation}
< PQii_3y \vert PQ i'i_3'y'> = \delta_{ii'} \delta_{i_3i_3'} \delta_{y
y'}~~.
\end{equation}
\indent The orthonormal basis in $V(P,0)$ can be constructed by the
normalization of Eq. (76) with the factor $a^P(i,i_3,y)$, such that
\begin{equation}
\vert P0ii_3y> = a^P(i,i_3,y) 
(x_1)^{p_1}(x_2)^{p_2}(x_3)^{p_3}~~.
\end{equation}
By acting with the operators $I^2$, $I_3$, $Y$ on both sides of this
equation one obtains 
\begin{equation}
p_1=i+i_3~~,~~p_2=i-i_3~~,~~p_3= \frac{P}{3} -y~~,
\end{equation}
and then, by using Eq. (77), one obtains a linear relationship between
isospin and hypercharge,
\begin{equation}
i=\frac{P}{3} + \frac{y}{2}~~.
\end{equation}
\indent The weights of the basis vectors of the space $V(P,0)$ are not
degenerate, and the states labeling can be achieved by using only 
$ ( i_3,y ) \equiv m $. Therefore, in the following will be used the 
notation
%eq98
\begin{equation}
\xi^m_P \equiv \xi^y_{P i_3}  \equiv \vert P0ii_3y> = 
a^P(i_3,y)  (x_1)^{p_1}(x_2)^{p_2}(x_3)^{p_3}~~.
\end{equation}
The complex conjugation of this equation leads to the expression of the 
contravariant basis vectors,
%eq99
\begin{equation}
\xi^Q_{-m} = (\xi^m_Q)^* = a^Q (i_3,y) (y^1)^{i+i_3} (y^2)^{i-i_3} (y^3)^{
\frac{Q}{3} -y}~~.
\end{equation}
The vector space generated by the basis $\xi^Q_m$ carries an irrep
equivalent to $D(0,Q)$, but where the matrix elements of the operators
$I_{\pm}$ and $K_{\pm}$ are not positive definite. The transition to a
basis $\eta^m_Q$ in which the phase convention is fulfilled, is generated
by the transformation of Eq.(67), such that
%eq100
\begin{equation}
\eta^{-m}_Q = a^Q (i_3,y) (-y^1)^{i+i_3} (y^2)^{i-i_3}
(y^3)^{\frac{Q}{3}-y}= (-1)^{i+i_3} (\xi^m_Q)^*~~.
\end{equation}
By using Eq. (89) and (63) it follows that Eq. (100) represents the canonical
basis of the space $V(0,Q)$,
%eq101
\begin{equation}
\eta_Q^{-m} \equiv \vert 0 Q i -i_3 -y> = (-1)^{ i+i_3} \vert Q0ii_3
y>^*~~,
\end{equation}
or, by using the notation 
%eq102
\begin{equation}
k=\frac{Q}{3} - \frac{y}{2}~~,
\end{equation}
%eq103
\begin{equation}
\eta^m_Q \equiv \eta^y_{Q k_3} = \vert 0 Q k k_3 y> = 
\end{equation}
$$
= (-1)^{\frac{Q}{3}- 
(k+ \frac{y}{2}) } a^Q ( -k_3,-y) (y^1)^{k-k_3} (y^2)^{k+k_3} (y^3)^{
\frac{Q}{3} +y} ~.
$$
{\bf VI. The Clebsch-Gordan series} \\[.5cm]
\indent The space obtained by the direct product of two irrep spaces 
 is not irreducible, and it should be decomposed in a direct sum of
irreducible spaces. The basis of the product space can be related to the
canonical basis of the direct sum of irrep spaces by a unitary
transformation $S$, such that
%eq104
\begin{equation}
S(V(P1,Q1) \otimes V(P2,Q2)) = \oplus_k \oplus_{ \gamma=1}^{m_k} V^\gamma
(P_k,Q_k)~~.
\end{equation}
This transformation brings the matrices of representation to a
block-diagonal form,  
%eq105
\begin{equation}
S(R(P1,Q1) \otimes R(P2,Q2) )) S^{-1} = \oplus_k m_k R(P_k,Q_k)~.
\end{equation}
The elements of the matrix $S$ are called Clebsch-Gordan coefficients, and
the set of irrep spaces $V^\gamma(P_k,Q_k)$ which appear in the right-hand
side of Eq. (104) represent the Clebsch-Gordan series. The index $\gamma$
is necessary to distinguish between the spaces which are isomorphic, and
carry the same irrep. The main difficulty which appears with respect to
the case of the group $SU(2)$ is due to the fact that in general, for
certain representations $D(P_k,Q_k)$,  $m_k$
can be greater than 1.  \\ \indent
The basis of the direct product space has the form of a mixed
tensor, without permutation symmetry. The linear combinations of the
components of this tensor obtained by contraction and
antisymmetrization generate subspaces which are $SU(3)$ invariant.
Therefore, the basis of the irrep spaces is generated by those mixed
tensors which are symmetric at the permutation of the upper and lower
indices, and traceless with respect to any contraction. The tensors which 
give the decomposition of the direct product can be constructed
explicitly below, for each particular case. \\[.5cm]
(A). $D(P_1,0) \otimes D(P_2,0)$ and $D(0,Q_1) \otimes D(0,Q_2)$
\\
The direct product basis is
%eq106
\begin{equation}
T_{(i_1...i_{P_1})(i_{P_1+1}...i_{P_1+P_2})} = T_{1i_1...i_{P_1}}
T_{2 i_{P_1+1} ...i_{P_1+P_2}}~~,
\end{equation}
a tensor symmetric to permutations within each subset of lower
indices. By contraction with the antisymmetric invariant contravariant
tensor $\epsilon^{ijk}$ one obtains the sequence of tensors $T^{(k)}$,
%eq107
\begin{equation}
T^{(k)j_1...j_k}_{(i_{k+1}...i_{P_1})(i_{P_1+k+1}...i_{P_1+P_2})}=
\epsilon^{j_k i_k i_{P_1+k}}   
T^{(k-1)j_1...j_{k-1}}_{(i_{k}...i_{P_1})(i_{P_1+k}...i_{P_1+P_2})}~,
\end{equation}
which are traceless with respect to every contraction, but  not 
symmetric with respect to the permutation of the lower indices. The spaces
$V^{(k)}$ generated by them satisfy the relationship
%eq108
\begin{equation}
V(P_1,0) \otimes V(P_2,0) = V^{(0)} \supset V^{(1)} \supset ...
\supset V^{(n)}~, n=\min(P_1,P_2)~.
\end{equation}
The tensors $T^{(k)}$ can be symmetrized with respect to every pair of
lower indices placed in different subsets. This procedure leads to the
irreducible tensors $T^{(k)}_S$,
%eq109
\begin{equation}
T^{(k)j_1...j_k}_{S (i_1...i_{P_1+P_2-2k})}=
T^{(k)j_1...j_{k}}_{(...i_r...)(...i_s...)}+
T^{(k)j_1...j_{k}}_{(...i_s...)(...i_r...)}~,
\end{equation}
which generate spaces denoted $V^{(k)}_S$.  The equations (107) and (109)
lead to a decomposition of the space $V^{(k)}$ of the form 
%eq110
\begin{equation}
V^{(k)}= V^{(k)}_S \oplus V^{(k+1)}~~,~~ V^{(k)}_S=V(P_1+P_2-2k,k)~~.
\end{equation}
By using the Eq. (108) and (110) one obtains the Clebsch-Gordan series
%eq111
\begin{equation}
D(P_1,0) \otimes D(P_2,0) = \sum_{k=0}^n D(P_1+P_2-2k,k) ~~,~~
n= \min(P_1,P_2)~~.
\end{equation} 
Similarly it is possible to show that 
%eq112
\begin{equation}
D(0,Q_1) \otimes D(0,Q_2) = \sum_{k=0}^n D(k,Q_1+Q_2-2k) ~~,~~
n= \min(Q_1,Q_2)~~.
\end{equation} 
(B).  $D(P,0) \otimes D(0,Q)$ \\
According to Eq. (87), the decomposition of this direct product has
the form
%eq113
\begin{equation}  
D(P,0) \otimes D(0,Q) = \sum_{k=0}^n D(P-k,Q-k)~~,~~ n= \min(P,Q)
\end{equation}
The basis of the space $V^{(k)}_0$ is represented by the traceless tensors
$T^{(k)}_0$, and can be constructed by using the method of projection
operators \cite{1} ch. 7 sect. 3. Thus,
%eq114
\begin{equation}
T^{(k)~ j_{k+1}...j_Q}_{0~~~i_{k+1}...i_P}=
\oint_{SU(3)} dU [R(P-k,Q-k)_U]_{ab} 
U^{\beta_{k+1}}_{i_{k+1}} ... 
\end{equation}
$$
... U^{\beta_P}_{i_P} 
(U^{\alpha_{k+1}}_{j_{k+1}} )^*...
(U^{\alpha_Q}_{j_Q})^*
T^{(k)~\alpha_{k+1}... \alpha_Q}_{~~ \beta_{k+1}...
\beta_P} ~~. 
$$
Here $U \in SU(3)$ denotes a group element, $dU$ is an invariant
measure on $SU(3)$,  and $[R(P-k,Q-k)_U]_{ab}$ is a matrix element of
$U$ in the representation $D(P-k,Q-k)$.
The indices $a,b$, in the vector notation, are such that
$a \equiv(i,i_3,y)$ it is the same as $(^{j_{k+1}...j_Q}_{i_{k+1}...i_P})$
in the tensor notation, and $b$ is arbitrary, but fixed.   \\[.5cm]
(C).  $D(P_1,Q_1) \otimes D(P_2,Q_2)$. \\
The basis of the reducible space obtained by the direct product is
represented by the components of the tensor
%eq115
\begin{equation}
T^{(j_1...j_{Q_1})(j_{Q_1+1} ...j_{Q_1+Q_2})}_{(i_1...i_{P_1})(i_{P_1+1}
...i_{P_1+P_2})} =
T^{~~j_1...j_{Q_1}}_{1~i_1...i_{P_1}}
T^{~~j_{Q_1+1} ...j_{Q_1+Q_2}}_{2~i_{P_1+1}
...i_{P_1+P_2}}~~,
\end{equation}
where $T_1$ and $T_2$ are traceless symmetric tensors. The reduction can
be performed in this case by using the procedure suggested by Coleman in
ref. \cite{30}, consisting of two steps:  \\
(1) - the product space is
decomposed in a direct sum of reducible invariant spaces 
$V(P,P'; Q,Q')$ generated by traceless tensors. \\
(2) - the spaces $V(P,P';Q,Q')$ are decomposed in a sum of irreducible 
spaces. \\[.5cm] 
(1) - Following the procedure applied in the case (B), it is possible to
construct traceless tensors  $T^{(m,n)}$   
%eq116
\begin{equation}
T^{(m,n)~(j_{n+1}...j_{Q_1})(j_{Q_1+m+1}
...j_{Q_1+Q_2})}_{~~~~~(i_{m+1}...i_{P_1})(i_{P_1+n+1}
...i_{P_1+P_2})} = 
\delta^{i_m}_{j_{Q_1+m}}
T^{(m-1,n)~(j_{n+1}...j_{Q_1})(j_{Q_1+m}
...j_{Q_1+Q_2})}_{~~~~~(i_m...i_{P_1})(i_{P_1+n+1}
...i_{P_1+P_2})} =
\end{equation}
$$
\delta^{i_{P_1+n}}_{j_n}
T^{(m,n-1)~(j_{n}...j_{Q_1})(j_{Q_1+m+1}
...j_{Q_1+Q_2})}_{~~~~~(i_{m+1}...i_{P_1})(i_{P_1+n}
...i_{P_1+P_2})}~~.
$$
The spaces generated by the tensors $T^{(m,n)}$ are denoted $V^{(m,n)}$,
and the orthogonal complements of the spaces $V^{(m+1,n)}$ and
$V^{(m,n+1)}$ in $V^{(m,n)}$ are denoted by $V^{(m_0,n)}$ and 
$V^{(m,n_0)}$, respectively. The bases of these spaces can be obtained,
for instance, by using Eq. (114), and 
$V^{(m_0,n_0)} = V(P_1-m,P_2-n; Q_1-n,Q_2-m)$. Because
%eq117
\begin{equation}
V^{(m,n)} = V^{(m_0,n_0)} \oplus V^{(m_0,n+1)} \oplus V^{(m+1,n_0)}
\oplus V^{(m+1,n+1)}
\end{equation}
one obtains the decomposition
%eq118
\begin{equation}
V(P_1,Q_1) \otimes V(P_2,Q_2) = \sum_{m_0=0}^{ \min(P_1,Q_2)} 
\sum_{n_0=0}^{ \min(P_2,Q_1)} V^{(m_0,n_0)} = 
\end{equation}
$$
=  \sum_{m=0}^{ \min(P_1,Q_2)} 
\sum_{n=0}^{ \min(P_2,Q_1)} V(P_1-m,P_2-n;Q_1-n,Q_2-m) ~~.
$$
(2) - Let  
$T^{(j_1...j_s)(j_{s+1}
...j_{s+s'})}_{0~~(i_{1}...i_r)(i_{r+1} ...i_{r+r'})} 
\equiv T^{~~(s)(s')}_{0~(r)(r')}$ a traceless basis tensor
in $V(r,r';s,s')$. By symmetrization with respect to any pair of upper 
indices belonging to different subsets one obtains a tensor 
$T^{~~(s+s')}_{0~ (r)(r')}$. This is symmetric in the upper indices, and 
represents the basis of a space denoted $V(r,r';s+s')$. The orthogonal 
complement of the space $V(r,r';s+s')$ in $V(r,r';s,s')$, denoted
$V(r+r'+1;s-1,s'-1)$, is generated by the tensor
%eq119
\begin{equation} 
\epsilon_{i j_1j_{s+1}} T^{(j_1...j_s)(j_{s+1}
...j_{s+s'})}_{0~~(i_{1}...i_r)(i_{r+1} ...i_{r+r'})} 
\end{equation}
which is traceless and symmetric in all lower indices. Thus,
%eq120
\begin{equation}
V(r,r';s,s') = V(r,r';s+s') \oplus V(r+r'+1;s-1,s'-1)~~.
\end{equation}
The complete decomposition of the subspaces $V(r,r';s+s')$ and
$V(r+r'+1;s-1,s'-1)$ can be obtained by the method used in the case (A),
and the corresponding series are similar to Eq. (111) and
(112). Therefore, the final result is
%eq121
\begin{equation}
V(r,r';s,s') = V(r+r';s+s') \oplus 
\sum_{k=1}^{ \min(r,r')} V(r+r'-2k;s+s'+k) 
\end{equation}
$$
\oplus \sum_{k=1}^{\min(s,s')} V(r+r'+k,s+s'-2k)~~.
$$
\indent
The formulas of Eq. (118) and (121) solve the problem of the
Clebsch-Gordan series for $SU(3)$ in the general case. The same 
permutation symmetry arguments as above, lead to the rules which give 
the decomposition of the product between the Young tableaux associated to
the irreps \cite{1} ch. 8. The problem of the decomposition of the direct
product for an arbitrary  semisimple Lie group was solved by Kostant and
Steinberg \cite{1} ch. 8 sect. 8. A purely geometrical procedure to obtain
the decomposition was given by Speiser \cite{9}, \cite{31}.        
\\[.5cm]
{\bf VII. The weights multiplicity } \\[.5cm] \indent
The explicit form of the canonical basis of the spaces $V(P,0)$ and
$V(0,Q)$ is given by Eq. (98) and (103), respectively. The canonical basis 
of $V(P,Q)$ can be obtained by decomposing the direct product between  
$V(P,0)$ and $V(0,Q)$. By using the notation $s \equiv (P,Q)$ for the
irreps labels, Eq. (104) leads to 
%eq122
\begin{equation}
\vert s i i_3 y>_\gamma = \sum_{ \mu m j k} 
\end{equation}
$$
<^{ s_1~~~~~~~s_2 ~~~~~}_{jm
\mu ~k ~i_3-m ~y - \mu} \vert^{ ~ s \gamma~}_{i i_3 y} > 
\vert s_1 j m \mu> \vert s_2 k ~i_3-m ~ y- \mu >  
$$
where  
$<^{ s_1~~~~~~~s_2 ~~~~~}_{jm
\mu ~k ~i_3-m ~y - \mu} \vert^{ ~ s \gamma~}_{i i_3 y} > $ are the $SU(3)$ 
Clebsch-Gordan coefficients, and $\gamma$ is a label used to distinguish
between the subspaces which all carry the same irrep $s$. Because 
%eq123
\begin{equation}
\vert s_1 j m \mu> \otimes \vert s_2 k ~i_3-m ~y- \mu >  
\end{equation}
are not eigenstates of the operator $I^2$, and $SU(3) \supset U(1) \otimes
SU(2)$, it is convenient to consider the transformation matrix $S$ as a
product of two unitary matrices, denoted $\alpha$ and $C$, such
that
%eq124
\begin{equation}
S = \alpha C
\end{equation}
and
%eq125
\begin{equation}
\alpha \alpha^\dagger = I~~,~~ CC^\dagger = I~~.
\end{equation}
The elements of the matrices $C$ and $\alpha$ are the $SU(2)$
Clebsch-Gordan coefficients, and, respectively, the isoscalar factors. By
using this factorization (the Racah lema \cite{32}), Eq. (122) becomes  
%eq126
\begin{equation}
\vert s i i_3 y>_\gamma = \sum_{ \mu m j k} 
\alpha^{ i y s s_1 s_2 \gamma}_{\mu ~ j ~ k} \times
\end{equation}
$$
C^{ j ~~k ~~i}_{m~ i_3-m ~ i_3} 
\vert s_1 j m \mu> \vert s_2 k ~i_3-m ~y- \mu >  
$$
with
%eq127
\begin{equation}
C^{ j ~k~i}_{j_3 k_3 i_3} \equiv ( < jj_3 \vert \otimes < k k_3 \vert ) 
\vert jk i i_3 >~~. 
\end{equation} 
If $s_1=(P,0)$, $s_2=(0,Q)$ and $s=(P,Q)$, then the summation on $j$ and
$k$ in Eq. (126) reduces to the one on $\mu$, and the expression of the
canonical basis in the space $V(P,Q)$ becomes
%eq128
\begin{equation}
\vert PQ i i_3 y> = \sum_{ \mu m } 
\alpha^{ i y~ (PQ)~ (P0)~ (0Q) }_{\mu ~ \frac{P}{3}+ \frac{ \mu}{2}  ~ 
\frac{Q}{3} - \frac{ y - \mu}{2} }  
C^{ \frac{P}{3} + \frac{ \mu}{2}  ~ \frac{Q}{3} - \frac{ y- \mu}{2}  
~ i}_{m ~~~ i_3-m ~~~ i_3} \xi^\mu_{Pm} \eta^{ y - \mu}_{Q i_3-m} ~~.
\end{equation}
The weights multiplicity is given by the number of the values taken by $i$
in this expression, at fixed values of $i_3$ and $y$. By using the
properties of the coefficients $C$ one obtains 
%eq129
\begin{equation}
\vert k-j \vert \le i \le k+j~~,
\end{equation}
where $j= P/3+ \mu/2$ and $k=Q/3-(y- \mu)/2$. \\ \indent
The minimum value of $i$ is the same for all the terms of the sum in
Eq. (128), equal to 
%eq130
\begin{equation}
i_{min} = \vert k-j \vert = \vert \frac{Q-P}{3} - \frac{ y}{2} \vert=
\end{equation}
$$
\frac{P-Q}{3} + \frac{y}{2} ~~if~~y \ge - \frac{2}{3} (P-Q)~~~~~(a)
$$
and 
$$
\frac{Q-P}{3} - \frac{y}{2} ~~if~~y \le - \frac{2}{3} (P-Q)~~~~~(b)~~.
$$
The maximum value of $i$ for an arbitrary term of the sum is 
$$
i_{max} = k+j= \frac{P+Q}{3} - \frac{y}{2} + \mu~~,
$$        
different from term to term. The highest value of $i_{max}$ will be
denoted by $i_M$. The coefficients $C^{j~k~i}_{j_3 k_3 i_3}$ vanish when
$i > j+k$, but the sum of Eq. (128) is not zero if it contains at least
one term, which means $i \le i_M$. The range of $i$ in $\vert PQ ii_3 y>$
is limited by the inequalities
%eq131
\begin{equation}
\vert i_3 \vert \le i ~~,~~ i_{min} \le i \le i_M
\end{equation}
where
%eq132
\begin{equation}
i_M = \frac{P+Q}{3} - \frac{ y}{2} + ( \mu)_{max}~~.
\end{equation}
The maximum value of $\mu$ which appears in Eq. (128), denoted $(
\mu)_{max}$, can be found by using Eq. (77), (79), (96) and (103),   
%eq133
\begin{equation}
p_3= \frac{P}{3} - \mu~~,~~ q_3 = \frac{Q}{3} + y - \mu~~,
\end{equation}
%eq134
\begin{equation}
0 \le p_3 \le P~~,~~ 0 \le q_3 \le Q~~.
\end{equation}
These lead to the inequalities
%eq135
\begin{equation}
- \frac{2}{3} P \le \mu \le \frac{P}{3} ~~,~~ y - \frac{2}{3} Q \le \mu 
\le \mu + \frac{Q}{3}~~,
\end{equation}
which give the maximum value 
%eq136
\begin{equation}
( \mu)_{max} = \min( \frac{P}{3} , y+ \frac{Q}{3} ) = 
\end{equation}
$$ 
\frac{P}{3} ~~if~~ y \ge \frac{ P-Q}{3} 
$$ 
and
$$
y+ \frac{Q}{3} ~~if~~ y \le \frac{P-Q}{3}~~.
$$
This result shows that the value of $i_M$ is
%eq137
\begin{equation}
i_M = \frac{2P+Q}{3} - \frac{y}{2} ~~if~~ y \ge \frac{ P-Q}{3} ~~(a)~,
\end{equation}
and  
$$
i_M = \frac{P+2 Q}{3} + \frac{y}{2} ~~if~~ y \le \frac{ P-Q}{3} ~~(b)~.
$$
The equations (130) and (137) can be represented in the orthogonal frame
$(i,y)$ by straight lines. Thus, one obtains the diagram of Fig. (3),
bounded by the lines $u$, $d$, $\nu$ and $\delta$ defined by the equations
$$
u:~~ i= \frac{y}{2} + \frac{P+2Q}{3} ~~,~~ \nu :~~ i = \frac{y}{2} +
\frac{P-Q}{3} ~~,
$$ 
$$
d:~~ i= - \frac{y}{2} + \frac{2 P+Q}{3} ~~,~~ \delta :~~ i = - \frac{y}{2}
- \frac{P-Q}{3} ~~. 
$$ 
These lines cross at the points $A$, $B$, $C$, $D$, with coordinates
$$
A( \frac{Q}{2}, - \frac{2 P+Q}{3} )~~;~~
B( \frac{P+Q}{2}, \frac{P-Q}{3} )~~;~~
$$
$$
C( \frac{P}{2}, \frac{P+2 Q}{3} )~~;~~
D( 0 , \frac{2}{3}(Q-P) )~~.
$$ 
The quantum numbers $(i, y)$ in Eq. (128) may take values only within the 
parallelogram bounded by these lines. The whole set of quantum numbers
$(i,i_3,y)$ which label the canonical basis of $V(P,Q)$ can be represented
by the sites of a 3-dimensional lattice (Fig. 4). The projection of this
lattice in the plane $(i_3,y)$ is the weight diagram of Fig. 2, and its 
volume is equal to the dimension of the space $V(P,Q)$, given in 
Eq. (88) \cite{33}. The same result was obtained by Biedenharn \cite{16}
-II by using the matrix elements of the operators $K_{\pm}$ and
$L_{\pm}$. The formulas which allow the calculus of the weights 
multiplicities for any semisimple Lie group have been given by Kostant
(1959) and Freudenthal (1969) \cite{23} ch. X sect. 13.4. \\[.5cm]
{\bf VIII. The matrix elements of the generators} \\[.5cm]
\indent The $U(n)$ irrep spaces and the matrix elements of the
generators can be obtained by using the Gel'fand-Zetlin method \cite{1}
ch. 10. In this work the matrix elements of the operators $I_{\pm}$,
$K_{\pm}$ and $L_{\pm}$ are calculated by extending the procedure applied
for $SU(2)$, of using the matrix elements of the commutators
$[A^i_k,A^k_i]$ which are known. \\ \indent
(A). Denoting by $B_{ii_3}$ the matrix elements of the operator $I_-$,
the commutation relations 
%eq138
\begin{equation}
[Y,I_{\pm}]=0~~,~~[I_3,I_\pm]= \pm I_\pm~~,~~[I^2,I_\pm]=0
\end{equation}
show that
%eq139
\begin{equation}
I_- \vert PQii_3y> = B_{ii_3} \vert PQi ~i_3-1~ y>~~,~~
\end{equation}
$$
I_+ \vert PQii_3y> = B_{ii_3+1} \vert PQi ~i_3+1~ y>~~.
$$  
Thus, the matrix element of the commutator $[I_+, I_-]= 2 I_3$ in the
state $\vert PQii_3y>$ is
%eq140
\begin{equation}
B^2_{ii_3} - B^2_{ii_3+1} = 2 i_3~~.
\end{equation}
By using the conditions $B_{ii_3} \ge 0$, $B_{i-i}=0$, one obtains
%eq141
\begin{equation}
B_{ii_3} = \sqrt{ i(i+1) - i_3 (i_3-1)}~~.
\end{equation}
\indent (B). In the adjoint representation (Eq. (19) (29)) the operators
$K_\pm$ and $L_\pm$ are eigenvectors of $ad_{I_3}$ and $ad_Y$,
%eq142
\begin{equation}
[I_3,K_\pm] = \frac{ \pm 1}{2} K_\pm~~,~~ [Y, K_\pm] = \pm K_\pm~~,
\end{equation} 
$$
[I_3,L_\pm] =- \frac{ \pm 1}{2} L_\pm~~,~~ [Y, L_\pm] = \pm L_\pm~~,
$$  
but not for $ad_{I^2}$, because
%eq143
\begin{equation}
[I^2,K_+] = L_+ I_++K_+ (\frac{3}{4} + I_3)~~,~~
[I^2,L_+] = K_+ I_-+L_+ (\frac{3}{4} - I_3)~~.
\end{equation}
Therefore, the vectors $K_+ \vert PQii_3y>$ and $L_+ \vert PQii_3 y>$ 
are linear combinations of the canonical basis vectors of $V(P,Q)$ which
have the weights given by Eq. (57),
%eq144
\begin{equation}
K_+ \vert PQ~i~i_3~ y> = \sum_{i'} 
\gamma^{i'}_{ii_3y} \vert PQ~i'~ i_3+ \frac{1}{2}~ y+1> ~~,
\end{equation}
%eq145
\begin{equation}
L_+ \vert PQ~i~i_3~ y> = \sum_{i'} \omega^{i'}_{ii_3y} \vert PQ~i'~ i_3-
\frac{1}{2}~ y+1> ~~.
\end{equation}
By acting with the operator $I^2$ on both sides of Eq. (144) and (145),
and then by using Eq. (141), (143) and (94), we obtain a homogeneous
system of equations for the coefficients $\gamma^i_{ii_3y}$ and
$\omega^{i'}_{ii_3y}$. The determinant of this system vanishes only  if
%eq146
\begin{equation}
i' = i \pm \frac{1}{2}~~,
\end{equation}  
and in this case the solution is
%eq147
\begin{equation}
\gamma^{i + \frac{1}{2}}_{i~i_3~y} = \sqrt{ \frac{i+i_3+1}{i-i_3} }
\omega^{i+ \frac{1}{2}}_{i~ i_3+1~ y}~~,
\end{equation}
$$
\gamma^{i - \frac{1}{2}}_{i~i_3~y} = - \sqrt{ \frac{i-i_3}{i+i_3+1} }
\omega^{i- \frac{1}{2}}_{i~ i_3+1~ y}~~.
$$
According to Eq. (146) the sums in Eq. (144) and (145) contain only two
terms, such that
%eq148
\begin{equation}
K_+ \vert PQ~i~i_3~y>=  
\gamma^{i+ \frac{1}{2} }_{ii_3y} \vert PQ~i+ \frac{1}{2}~ i_3+
\frac{1}{2}~ y+1> +
\end{equation}
$$
\gamma^{i- \frac{1}{2} }_{ii_3y} \vert PQ~i- \frac{1}{2} ~ i_3+
\frac{1}{2}~ y+1> ~~,
$$ 
%eq149
\begin{equation}
L_+ \vert PQ~i~i_3~y>=  
\omega^{i+ \frac{1}{2} }_{ii_3y} \vert PQ~i+ \frac{1}{2}~ i_3-
\frac{1}{2}~ y+1> +
\end{equation}
$$
\omega^{i- \frac{1}{2} }_{ii_3y} \vert PQ~i- \frac{1}{2} ~ i_3-
\frac{1}{2}~ y+1> ~~.
$$ 
The operator $K_+$ in Eq. (148) can be replaced by the commutator
$[I_+, L_+] = K_+$, and then, by using Eq. (149), (141), and (94), it is
possible to find two recurrence relationships in $i_3$ for $\gamma^{i +
\frac{1}{2}}_{ii_3y}$ and $\gamma^{i- \frac{1}{2}}_{ii_3y}$. 
Denoting by $\chi_{iy} \equiv \gamma^{i+ \frac{1}{2}}_{iiy}$ and
$\kappa_{iy} \equiv \gamma^{ i - \frac{1}{2}}_{i~-i~y}$, the solution of
these relationships is
%eq150
\begin{equation}
\gamma^{i+ \frac{1}{2}}_{ii_3y} = \sqrt{  \frac{ i+i_3+1}{2i+1}}
\chi_{iy} ~~,~~
\gamma^{i- \frac{1}{2}}_{ii_3y} = \sqrt{  \frac{ i-i_3}{2i}}
\kappa_{iy} ~~.
\end{equation}
This result, together with Eq. (147), determine the isospin dependent
factor of the coefficients $\omega$, as
%eq151
\begin{equation}
\omega^{i+ \frac{1}{2}}_{ii_3y} = \sqrt{  \frac{ i-i_3+1}{2i+1}}
\chi_{iy} ~~,~~
\omega^{i- \frac{1}{2}}_{ii_3y} = - \sqrt{  \frac{ i+i_3}{2i}}
\kappa_{iy} ~~.
\end{equation}   
By convention the matrix elements of the operators $\{ A^i_k; i,k=1,2,3
\}$ are real. Moreover, they should be positive for $I_\pm$ and $K_\pm$, 
which means that the coefficients $\chi$ and $\kappa$ are positive real
numbers. This property indicates that the operators $K_- = (K_+)^\dagger$
and $L_-= (L_+)^\dagger$ act on the canonical basis vectors according to
%eq152
\begin{equation}
K_- \vert PQ~i~i_3~y>=  
\gamma^{i }_{i+ \frac{1}{2}~i_3- \frac{1}{2} ~y-1} \vert PQ~i+
\frac{1}{2}~ i_3- \frac{1}{2}~ y-1> +
\end{equation}
$$
\gamma^{i }_{i- \frac{1}{2} ~ i_3 - \frac{1}{2} ~y-1} \vert PQ~i-
\frac{1}{2} ~ i_3- \frac{1}{2}~ y-1> ~~,
$$ 
%eq153
\begin{equation}
L_- \vert PQ~i~i_3~y>=  
\omega^{i }_{i+\frac{1}{2} ~ i_3+ \frac{1}{2} ~ y-1} \vert PQ~i+
\frac{1}{2}~ i_3+ \frac{1}{2}~ y-1> +
\end{equation}
$$
\omega^{i}_{i- \frac{1}{2} ~ i_3 + \frac{1}{2} ~ y-1 } \vert PQ~i-
\frac{1}{2} ~ i_3+ \frac{1}{2}~ y-1> ~~.
$$  
To determine the factors $\chi$ and $\kappa$ it is necessary to find two
new recurrence formulas, together with the related initialization
values. These formulas are provided by the matrix elements of the 
commutators $[K_+, K_-]$ and $[L_+, L_-]$ in an arbitrary state $
\vert PQ ~i~i_3~y>$,
%eq154
\begin{equation}
< PQ~i~i_3~y \vert [K_+, K_-] \vert PQ~i~i_3~y> = i_3 + \frac{3}{2} y~~,
\end{equation}
and
%eq155
\begin{equation}
< PQ~i~i_3~y \vert L_+, L_-] \vert PQ~i~i_3~y> = - i_3 + \frac{3}{2} y~~.
\end{equation}
By using Eq. (148), (149), (152), (153) to express the action of the
commutators in terms of $\gamma$ and $\omega$, and then Eq. (150), (151)
to express $\gamma$ and $\omega$ by $\chi$ and $\kappa$, we get
%eq156
\begin{equation}
\frac{ 1}{2i+1} \kappa^2_{i+ \frac{1}{2}~y-1} + 
\frac{ 1}{2(i+1)} \chi^2_{i- \frac{1}{2}~y-1} -
\frac{ 1}{2i+1} \chi^2_{i~y} -
\end{equation}
$$
\frac{ 1}{2(i+1)} \kappa^2_{i~y} =
\frac{ 3y}{2(i+1)}
$$
and
%eq157
\begin{equation}
- \frac{ 1}{2i+1} \kappa^2_{i+ \frac{1}{2}~y-1} + 
\frac{ 1}{2i} \chi^2_{i- \frac{1}{2}~y-1} - 
\frac{ 1}{2i+1} \chi^2_{i~y} +
\frac{ 1}{2i} \kappa^2_{i~y} =1~~.
\end{equation}
The equations (156) and (157) relate the factors $\chi$ at the points
$4:(i,y)$ and $3:(i- 1/2,y-1)$, to the factors $\kappa$ at the points
$4:(i,y)$ and $1: (i+1/2,y-1)$ (Fig. 5). By adding these equations
$\kappa^2_{i+\frac{1}{2}~y-1}$ is eliminated, and one obtains a
relationship between $\chi_3$, $\chi_4$ and
$\kappa_4$ (Fig. 5). This gives $\kappa^2_{i~y}$ as a function
of $\chi_3$ and $\chi_4$.  By the transformation $i \rightarrow 
i + \frac{1}{2}$, $y \rightarrow y-1$, one can further obtain
$\kappa^2_{i + \frac{1}{2} ~ y-1} $ as a function of  $\chi_1$ and
$\chi_2$. When the resulting expressions of $\kappa^2_{i~y}$ and
$\kappa^2_{i+ \frac{1}{2}~y-1}$ are introduced in Eq. (157), one obtains
%eq158
\begin{equation}
\frac{i+1}{i+ \frac{1}{2}} \chi^2_{i~y-1} + \chi^2_{i~y+1} =
\frac{i+ \frac{3}{2} }{i+ 1} \chi^2_{i + \frac{1}{2} ~y} +
\chi^2_{i- \frac{1}{2} ~ y} ~~.
\end{equation}
Similarly, one can find a homogeneous equation for $\kappa$,
%eq159
\begin{equation}
\kappa^2_{i~y-1} + \frac{ i + \frac{1}{2}}{i} \kappa^2_{i~y+1} = 
\frac{ i+1}{i+ \frac{1}{2}} \kappa^2_{i+ \frac{1}{2} ~y} +
\kappa^2_{i- \frac{1}{2}~ y} ~~.
\end{equation}
By introducing the notations
%eq160
\begin{equation}
\phi_{i~y} = \frac{i+1}{i+ \frac{1}{2}} \chi^2_{i~y-1} -
\chi^2_{i- \frac{1}{2} ~ y} ~~,
\end{equation}
and
$$
\psi_{i~y} = \frac{ i+1}{i+ \frac{1}{2}} \kappa^2_{i+ \frac{1}{2} ~y} -
\kappa^2_{i ~ y-1} ~~,
$$ 
one can see that Eq. (158) and (159) can be expressed as
%eq161
\begin{equation}
\phi_{iy} = \phi_{i+ \frac{1}{2} ~ y+1} ~~,
\end{equation}
respectively
%eq162
\begin{equation}
\psi_{iy} = \psi_{i- \frac{1}{2} ~ y+1}~~.
\end{equation}     
 These equations state the invariance of the function $\phi_{iy}$ on
lines parallel to $u$, and of the function $\psi_{iy}$ on lines
parallel to $d$. The action of the operators $A^i_k$ on the boundary 
lines of the diagram of Fig. 3 cannot lead to weights outside the diagram,
and therefore, the values of $\chi$ and $\kappa$ on the boundary will be
used to initiate the recurrence formulas. Denoting by $( \Delta, \tau)$ 
and $( \Omega, \theta)$ the coordinates $(i,y)$ of the points placed on
the line $d$, respectively $ \nu$, the initial conditions are 
%eq163
\begin{equation}
\chi_{ \Delta \tau} = 0~~,~~ \kappa_{ \Omega \theta} =0~~.
\end{equation}
These conditions indicate the fact that $\phi_{iy}$ and $\psi_{iy}$
vanish on $d$, respectively $\nu$, while Eq. (161), (162) show further
that they vanish everywhere. This result can be expressed in the form  
%eq164
\begin{equation}
z^2_{i~y-1} = z^2_{i- \frac{1}{2}~y} ~~,~~
w^2_{i+ \frac{1}{2} ~ y} = w^2_{i~ y-1} ~~,
\end{equation}
where the factors $z_{iy}$ and $w_{iy}$ are related to $\chi_{iy}$ and
$\kappa_{iy}$ by the relationships
%eq165
\begin{equation}
\chi^2_{iy} = \frac{ z^2_{iy} }{2(i+1)}~~,~~
\kappa^2_{iy} = \frac{ w^2_{iy}}{2i+1}~~,
\end{equation}
and satisfy the conditions
%eq166
\begin{equation}
z_{ \Delta \tau} = 0~~,~~ w_{ \Omega \theta} = 0~~.
\end{equation}
The factors $z$ and $w$ take the same values in all points of any line
parallel to $d$, respectively $u$. Therefore, to calculate $\chi$ and
$\kappa$ at any point of the diagram of Fig. 3 it is enough to know the
values of $z$ on $u$ or $\nu$, and of $w$ on $d$ or $\delta$. \\ \indent
When the expressions of $\chi$ and $\kappa$ given by Eq. (165)  are 
introduced in the system of Eq. (156), (157), one obtains a system 
in $z$ and $w$, which can be solved with respect to $z^2_{i- \frac{1}{2}~
y-1}$ and $w^2_{i+ \frac{1}{2}~y-1}$,
%eq167
\begin{equation}
\frac{1}{i} z^2_{ i - \frac{1}{2}~ y-1} = \frac{1}{i + \frac{1}{2}} 
z^2_{iy} - \frac{1}{i(2i+1)} w^2_{iy} + 3y+2i+2~~,
\end{equation}
%eq168
\begin{equation}
\frac{1}{i+1} w^2_{ i + \frac{1}{2}~ y-1} = \frac{1}{i + \frac{1}{2}} 
w^2_{i~y} + \frac{1}{(i+1)(2i+1)} z^2_{iy} + 3y-2i ~~.
\end{equation} 
By replacing $(i,y)$ with $( \Omega, \theta)$ in Eq. (167), respectively 
with $\Delta, \tau)$ in Eq. (168), and using Eq. (166), we obtain two
simple recurrence relationships between the factors $z$ on the line $\nu$,
and $w$ on $d$, 
%eq169
\begin{equation}
\frac{1}{ \Omega} z^2_{ \Omega - \frac{1}{2}~ \theta -1 }=
\frac{1}{ \Omega + \frac{1}{2}} z^2_{ \Omega  \theta} + 3 \theta +
2 \Omega +2~~,
\end{equation}
%eq170  
\begin{equation}
\frac{1}{ \Delta + 1} w^2_{ \Delta + \frac{1}{2}~ \tau -1 }=
\frac{1}{ \Delta + \frac{1}{2}} w^2_{ \Delta  \tau} + 3 \tau -
2 \Delta ~~.
\end{equation}
These recurrence formulas can be initialized by using the values
%eq171
\begin{equation}
z_{ \frac{P}{2} ~ \frac{P+2Q}{3}} =0~~,~~
w_{ \frac{P}{2} ~ \frac{P+2Q}{3}} =0~~,~~
\end{equation}
given by Eq. (166), and the solutions are
%eq172
\begin{equation}
z^2_{ \Omega \theta} = ( 2 \frac{P-Q}{3} + \theta +1) ( \frac{ P+2Q}{3}
-
\theta)( \frac{ 2P+Q}{3} + \theta + 2)
\end{equation}
%eq173
\begin{equation}
w^2_{ \Delta  \tau } = ( 2 \frac{2P+Q}{3} - \tau +1) ( \frac{ P+2Q}{3}
- \tau)( \frac{ Q-P}{3} + \tau + 1)~~.
\end{equation}
The values of $z^2$ and $w^2$ at any other point of the diagram of Fig. 3
can be obtained by using Eq. (164), such that
%eq174
\begin{equation}
z^2_{i~y} = ( \frac{ 2P+Q}{3} -i - \frac{y}{2}) ( \frac{P+2Q}{3} + i + 
\end{equation}
$$
\frac{y}{2}+ 2) ( \frac{P-Q}{3} + i + \frac{y}{2} + 1 )
$$   
%eq175
\begin{equation}
w^2_{i~y} = ( \frac{ Q-P}{3} + i - \frac{y}{2}) ( \frac{P+2Q}{3} -i +
\end{equation}
$$
\frac{y}{2} +1) ( \frac{2P+Q}{3} + i - \frac{y}{2} + 1 )~~.
$$     
The coefficients $\chi_{iy}$ and $\kappa_{iy}$ are further
determined by Eq. (165), and finally, Eq. (150), (151), complete the
calculation by giving the matrix elements of the generators.  
In the following the action of the generators $K_\pm$ and $L_\pm$  will be
expressed by using the notation of ref. \cite{9}, as
$$
K_\pm \vert PQ~i~i_3~y>=  
a_\pm ( i~i_3~y )  \vert PQ~i+ \frac{1}{2}~ i_3 \pm
\frac{1}{2}~ y \pm 1> +
$$
$$
b_\pm (i~i_3~y) \vert PQ~i- \frac{1}{2} ~ i_3 \pm
\frac{1}{2}~ y \pm 1> ~~,
$$ 
$$
L_\pm \vert PQ~i~i_3~y>=  
c_\pm (i~i_3~y) \vert PQ~i+ \frac{1}{2}~ i_3 \pm
\frac{-1}{2}~ y \pm 1> +
$$
$$
d_\pm (i~i_3~y) \vert PQ~i- \frac{1}{2} ~ i_3 \pm
\frac{- 1}{2}~ y \pm 1> ~~.
$$ 
The coefficients $a_\pm$, $b_\pm$, $c_\pm$, $d_\pm$,
can be found by the comparison between these equations and Eq. (148),
(149) (152), (153). \\ \indent
In the particular case $Q=0$ the diagram of Fig. 3 reduces to a line,
given by Eq. (97), and the matrix elements are
%eq176
\begin{equation}
a^P_+ (i_3, y) = \sqrt{ ( \frac{P}{3} + \frac{y}{2} + i_3 +1) (
\frac{P}{3} -y) } ~~,~~ b^P_+ = 0~~,
\end{equation}  
%eq177
\begin{equation}
c^P_+ (i_3, y) = \sqrt{ ( \frac{P}{3} + \frac{y}{2} - i_3 +1) (
\frac{P}{3} -y) } ~~,~~ d^P_+ = 0~~.
\end{equation}  
\indent In the case $P=0$ the diagram of Fig. 3 reduces to the line of
Eq. (102), and the matrix elements are
%eq178
\begin{equation}
a^Q_+=0~~,~~ b^Q_+ (i_3, y) = \sqrt{ ( \frac{Q}{3} - \frac{y}{2} - i_3) (
\frac{Q}{3} +y+1) } ~~,
\end{equation}  
%eq179
\begin{equation}
c^Q_+=0~~,~~ d^Q_+ (i_3, y) = - \sqrt{ ( \frac{Q}{3} - \frac{y}{2} +
i_3) ( \frac{Q}{3} + y +1) } ~~.
\end{equation}  
{\bf IX. The Clebsch-Gordan coefficients} \\[.5cm]  
\indent The Clebsch-Gordan series and the explicit form of the basis
tensors in the irrep spaces appearing in the decomposition of the direct
product $V(P_1,Q_1) \otimes V(P_2,Q_2)$ have been presented in sect. VI. 
The expressions obtained for these tensors become cumbersome
when the dimension of the factor spaces increases, and therefore, the 
tensor representation is useful only in simple cases. However, the 
canonical basis of the spaces $V^\gamma(P_k,Q_k)$ in Eq. (104) can be 
constructed by using the theorems {\bf I-III} presented in
sect. IV. According to these theorems, the basis can be constructed 
by following a three-step procedure: \\
1. Find the subspaces $W(P_k,Q_k)$ generated by the vectors given in
Eq. (122) which satisfy Eq. (48),(49), namely
%eq180
\begin{equation}
I_+ \vert P_k ~Q_k~ \underline{i} ~ \underline{i} ~
\underline{y} >_\gamma=0~~,~~
\end{equation}
%eq181
\begin{equation}
K_+ \vert P_k ~Q_k~ \underline{i} ~ \underline{i} ~ \underline{y}
>_\gamma=0~~,~~
L_+ \vert P_k ~Q_k~ \underline{i} ~ \underline{i} ~ \underline{y}
>_\gamma=0~~,
\end{equation}
and
%eq182
\begin{equation}
I_3 \vert P_k ~Q_k~ \underline{i} ~ \underline{i} ~ \underline{y}
>_\gamma= \underline{i}
\vert P_k ~Q_k~ \underline{i} ~ \underline{i} ~ \underline{y}
>_\gamma ~~,~~
\end{equation}
$$
Y \vert P_k ~Q_k~ \underline{i} ~ \underline{i} ~ \underline{y}
>_\gamma = \underline{y} \vert P_k ~Q_k~ \underline{i} ~ \underline{i} ~
\underline{y}
>_\gamma ~~,~~
$$
where $\gamma$ takes values from 1 to $m_k= \dim W(P_k,Q_k)$, $
\underline{i} = P_k/2$, $\underline{y} = (P_k+2Q_k)/3$, and
%eq183
\begin{equation}
A^i_k = A^{(1)i}_k \otimes I^{(2)} + I^{(1)} \otimes A^{(2)i}_k~~,
i,k=1,2,3~~.
\end{equation}
2. Chose an orthonormal basis in each space $W(P_k,Q_k)$, such that
\begin{equation}
%eq184
_\gamma< P_k ~Q_k~ \underline{i} ~ \underline{i} ~ \underline{y}
\vert P_k ~Q_k~ \underline{i} ~ \underline{i} ~ \underline{y}
>_{ \gamma '} = \delta_{ \gamma \gamma '}  ~~,~~ \gamma, \gamma ' =
1,m_k~~.
\end{equation}
3. Generate the basis of the subspaces $V^\gamma ( P_k,Q_k)$ by applying
the operators $I_-$, $K_-$, $L_-$ on the highest weight vectors 
$\vert P_k~ Q_k~ \underline{i} ~ \underline{i} ~ \underline{y} >_\gamma$. 
\\ \indent
If the tensor product is simply reducible, ( $m_k$ is 0 or 1), the 
decomposition is obtained simply by the transition to a basis of
$V(P_1,Q_1) \otimes V(P_2,Q_2)$ in which the Casimir operators constructed
with the generators of Eq. (183) are diagonal. Because $F$ and $G$ are
Hermitian, the subspaces corresponding to different eigenvalues $f$, $g$, 
and implicitly with different indices $(P,Q)$, are orthogonal and
irreducible. The matrix of the linear transformation from the 
eigenvectors of the set of operators 
$$
(a): \{ F_1,F_2,G_1,G_2,I^2,I_1^2, I_2^2,I_3,Y_1,Y_2   \}
$$
to the eigenvectors of the set
$$
(b): \{ F,G,F_1,F_2,G_1,G_2,I^2,I_3,Y \}
$$
is the matrix of the isoscalar factors, and it can be obtained by finding
the common eigenvectors of the set $(b)$. \\ \indent
The multiplicities appear when the space obtained by the direct product 
has a symmetry higher than $SU(3)$. This means that it exists a Hermitian
operator $X$ which commutes with all the operators of the set $(b)$, and
is independent of them, which should be added to obtain the complete set
of compatible observables. The expression given by Moshinsky for this 
operator is
%eq185
\begin{equation}
X = \frac{1}{2} \sum_{i,j,k=1}^3 ( A^{(1)i}_k A^{(1)j}_i A^{(2)k}_j+
A^{(1)i}_k A^{(1)k}_j A^{(2)j}_i )~~.
\end{equation}
The subspaces generated by the eigenvectors of the set (b) of operators
can be further decomposed by diagonalizing the operator $X$. This
method is cumbersome, and in practice it is not used, but is equivalent
to the three-step procedure presented above. Thus, every vector satisfying
the system of Eq. (180)-(182), is also an eigenvector of the operators
$F$ and $G$. This follows by using the commutation relations of Eq. (23)
to rearrange the operators appearing in Eq. (54) such that $I_+$, $K_+$
and $L_+$ are placed to the right of $I_-$, $K_-$, $L_-$. When the
resulting expression is applied on the vectors which satisfy the
Eq. (180)-(182),
the only non-vanishing terms will be those dependent only on 
$I_3$ and $Y$, which lead to multiplication by a constant. Because $F$ and
$G$ are Hermitian, the solutions of Eq. (180)-(182), denoted by $\vert
s>$, which correspond to different sets $(P,Q)$, will be orthogonal. The     
remaining eigenvectors are generated in the step 3. They will be
eigenvectors of $F$ and $G$ with the same values as $\vert s>$, because 
$F$ and $G$ commute with all $A^i_k$. This shows that the solutions $\vert
s>$ corresponding to different values $P$ and $Q$ are highest
weight vectors in orthogonal irreducible subspaces. \\ \indent
When there are  multiplicities, there are several independent  solutions,
$\{ \vert s>_\gamma; \gamma=1, \dim(W_s) \}$, with the same indices
$(P,Q)$. In this case, one of the vectors $\vert s>$ is  arbitrarily
chosen,
and then the other can be obtained by the Gram-Schmidt procedure, such that
are orthonormal. This leads to a decomposition of the
product space in a direct sum of irreducible orthogonal subspaces, and
to an unitary matrix of isoscalar factors. The vectors of the canonical 
basis in these subspaces are linear combinations of the form of Eq. (126)
%eq186        
\begin{equation}
\vert s ii_3y> = \sum_{ \mu,m,j,k} \alpha^{ i y s s_1s_2}_{ \mu jk} 
C^{j~~k~~i}_{m~i_3-m~i_3} \vert s_1 j m \mu> \vert s_2 k~ i_3-m~ y- \mu>
~~.
\end{equation}
\indent For given values of $s_1$ and $s_2$, $s$ in Eq. (186) can take all
the values appearing in the Clebsch-Gordan series. Fixing a certain value
of $s$ means to specify the allowed intervals for the variation of $i$ and
$y$. These intervals have been determined in sect. VII, and the result is
presented in the diagram of Fig. 3. \\ \indent
The possible values of the summation
indices $\{ j, \mu \}$ and $\{ k, y - \mu \}$ can be represented in
similar diagrams, denoted $A$ and $B$, corresponding to the carrier spaces
of the irreps $s_1$, respectively $s_2$. In the orthogonal frame of
axes $(k, \mu,j)$, the possible values of these indices are represented by
the points $\pi$ for which the projection in the plane $(j, \mu)$ is a
node of the lattice $A$, and in the plane $(k, \mu)$ a node of the
lattice $B$. This condition is expressed by the inequalities
%eq187
\begin{equation}
- \frac{ 2 P_1+Q_1}{3} \le \mu \le \frac{ P_1+ 2Q_1}{3}
\end{equation}
%eq188
\begin{equation}
- \frac{ 2 P_2+Q_2}{3} \le y- \mu \le \frac{ P_2+ 2Q_2}{3}~~,
\end{equation}
or, in compact form, 
%eq189
\begin{equation}
\max ( - \frac{ 2 P_1+Q_1}{3}, y 
- \frac{ P_2+2 Q_2 }{3} ) 
\end{equation}
$$
 \le \mu \le \min ( \frac{ P_1+ 2Q_1}{3},
y+ \frac{ 2 P_2 + Q_2 }{3}) ~~.
$$
However, not all the points $(k, \mu, j)$ which satisfy this
inequality appear as summation indices in Eq. (186), but only those 
for which 
%eq190
\begin{equation}
 k+j \ge i ~~,~~ \vert k - j \vert \le i~~.
\end{equation}
Geometrically, this means that not all the points selected by Eq. (189), 
represented by the 3D lattice of Fig. 6, appear as indices of the
isoscalar factors, but only those which satisfy also Eq. (190).     
This condition selects the points which are inside and on the border of
the space bounded by three planes which are  parallel to the
$\mu$-axis, and cross the $k$ and $j$ axes at the points with coordinates
$(k, j) =$ $(i,0)$, $(0,i)$, and $(0, -i)$ (Fig. 7). The final lattice
obtained by this selection procedure is complicated in general, and it
changes at the variation of $i$ and $y$. \\ \indent
With these considerations, we can proceed by following the
three-step procedure presented at the beginning of this section. \\
1. According to Eq. (186), the highest weight vectors have the expression 
%eq191
\begin{equation}
\vert s \underline{i} ~ \underline{ i } ~ \underline{y} > = \sum_{
\mu,m,j,k} \alpha^{s}_{ \mu jk} 
C^{j~~k~~ \underline{i} }_{m~ \underline{ i } -m~ \underline{ i} } \vert
s_1 ~ j ~ m ~ \mu> \vert s_2 ~ k ~ \underline{i}-m ~ y- \mu> ~~,
\end{equation}    
which satisfies Eq. (180) and (182) ($\alpha^s_{\mu j k} \equiv
\alpha^{ \underline{i} ~ \underline{y} ~ s s_1 s_2}_{\mu j k}$). By using
Eq. (181) with the notations of Eq. (150), one obtains the following
recurrence relations
%eq192
\begin{equation}
\sqrt{ \frac{ \underline{i} + j -k}{2j}  } \chi^{s_1}_{ j- \frac{1}{2} ~
\mu -1} \alpha^s_{ \mu-1 ~ j- \frac{1}{2} ~ k+ \frac{1}{2} } 
\end{equation}
$$
- \sqrt{ \frac{ (j+k- \underline{i} +1)( \underline{i} +k -j
+1) }{(2j+1)( \underline{i} + j + k +2) } } \kappa^{s_1}_{ j+ \frac{1}{2}
~ \mu -1} \alpha^s_{ \mu-1 ~ j+ \frac{1}{2} ~ k+ \frac{1}{2} } 
$$
$$
+ \sqrt{ \frac{ \underline{i} - j + k+1}{2k+1}  } \chi^{s_2}_{ k  ~
\underline{y} - \mu } \alpha^s_{ \mu j  k } 
$$
$$
+ \sqrt{ \frac{ (j+k- \underline{i} +1)( \underline{i} - k + j
) }{2(k+1)( \underline{i} + j + k +2) } } \kappa^{s_2}_{ k+ 1 
~ \underline{y} - \mu } \alpha^s_{ \mu ~ j ~ k+ 1} = 0
$$
and  
 %eq193
\begin{equation}
\sqrt{ \frac{ \underline{i} + j -k}{(2j)^2}  } \chi^{s_1}_{ j- \frac{1}{2}
~ \mu -1} \alpha^s_{ \mu-1 ~ j- \frac{1}{2} ~ k+ \frac{1}{2} } 
\end{equation}
$$
+ \sqrt{ \frac{ ( \underline{i} +k -j +1)( j+k - \underline{i} 
+1) }{(2j+1)( \underline{i} + j + k +2) } } \kappa^{s_2}_{ j+ \frac{1}{2}
~ \mu -1} \alpha^s_{ \mu-1 ~ j+ \frac{1}{2} ~ k+ \frac{1}{2} } 
$$
$$
- \sqrt{ \frac{( \underline{i} - j + k+1)(j+k- \underline{i})^2}{2j(2k+1)}
} \chi^{s_2}_{ k ~ y- \mu } \alpha^s_{ \mu  j  k } 
$$
$$
+ \sqrt{ \frac{ (\underline{i} +k -j +2)^2( \underline{i} - k + j
) (j+k- \underline{i} +1 }{4j(k+1)( \underline{i} + j + k +2) } }
\kappa^{s_2}_{ k+ 1 ~ \underline{y} - \mu } \alpha^s_{ \mu ~ j ~ k+ 1} =
0~~.
$$ 
These relationships contain the factors $\alpha$ at the points 
$A(k+1, \mu, j)$, $B(k, \mu, j)$, $C( k+ 1/2, \mu -1, j+ 1/2)$, and
$D(k+1/2, \mu -1, j- 1/2)$. By eliminating $\alpha_C \equiv \alpha^s_{ \mu
-1 ~ j+ \frac{1}{2} ~ k+ \frac{1}{2} } $ between them one obtains a
relationship between $ \alpha_A$, $\alpha_B$ and $\alpha_D$, 
%eq194
\begin{equation}
 \frac{ 2j+1}{ \sqrt{2j}} \chi^{ s_1}_{j - \frac{1}{2} ~ \mu -1} 
\alpha^s_{ \mu -1 ~ j - \frac{1}{2} ~ k + \frac{1}{2} } 
\end{equation}
$$
+ \sqrt{\frac{ ( \underline{i} + j -k)( \underline{i} + k -j+1) }{ 2k+1 }  
} \chi^{s_2}_{k ~ \underline{y} - \mu } \alpha^s_{ \mu ~ j ~ k} 
$$  
$$
+ \sqrt{\frac{ ( \underline{i} + j + k+2)( j+k-\underline{i} +1) }{
2(k+1) } } 
 \kappa^{s_2}_{k+1 ~ \underline{y} - \mu } \alpha^s_{ \mu ~ j ~ k+1}
=0~~,
$$  
and by eliminating $ \alpha_D \equiv \alpha^s_{ \mu -1 ~ j - 1/2 ~ k+ 1/2}
$,  a relationship between $ \alpha_A$, $\alpha_B$ and $\alpha_C$, 
%eq195
\begin{equation}
\sqrt{ \frac{ 2j+1}{ \underline{i} + j + k +2  } } \kappa^{ s_1}_{j +
\frac{1}{2} ~ \mu -1}  \alpha^s_{ \mu -1 ~ j + \frac{1}{2} ~ k +
\frac{1}{2} }  
- \sqrt{ \frac{j+k-  \underline{i} + 1 }{ 2k+1 }  
} \chi^{s_2}_{k ~ \underline{y} - \mu } \alpha^s_{ \mu  j  k} 
\end{equation}  
$$
+ \sqrt{\frac{ ( \underline{i} + j - k)( \underline{i}+k-j +1) }{
2(k+1)( \underline{i} +j+k+2) } } 
 \kappa^{s_2}_{k+1 ~ \underline{y} - \mu } \alpha^{s_2}_{ \mu ~ j ~ k+1}
=0~~.
$$  
These equations, combined with the inequalities
%eq196
\begin{equation}
\max ( - \frac{ 2 P_1+Q_1}{3},  
\frac{ P - P_2+2(Q- Q_2) }{3} )  
\end{equation}
$$
 \le \mu \le \min ( \frac{ P_1+ 2Q_1}{3},
\frac{ P + Q_2 + 2( P_2 +Q) }{3}) ~~,
$$
and
%eq197
\begin{equation}
\vert j -k \vert \le \frac{P}{2} \le j+k
\end{equation}
determine the isoscalar factors for the highest weight vectors. The
solution of these recurrence relationships is unique if all $\alpha^s_{
\mu jk}$ can be expressed in terms of only one of them. In this case,
the absolute value of this reference factor can be found by using the 
normalization relation
%eq198
\begin{equation}
\sum_{ \mu j k} \vert \alpha^s_{ \mu ~ j ~ k} \vert^2 =1~~,
\end{equation}
while its sign is given by the convention used in \cite{9} : among the
isoscalar factors of maximum $j$, are chosen as positive those of maximum
$k$. \\ \indent
The occurrence of the multiplicities is determined by the structure of the
recurrence relationships of Eq. (194) and (195), with the constraints of 
Eq. (196), (197). Let $\mu_{ min}$ and $\mu_{ max}$ be the effective
extreme values of $\mu$, different in general from the limits appearing in 
Eq. (196). One can see that the recurrence relationships determine
completely the factors $\alpha$ only in two cases:  \\
I. The lattice of Fig. 6 attached to the highest weight $( \underline{i},
y)$ is bounded at $\mu_{ min}$ by a segment parallel to the axis $k$, and
at $\mu_{max}$ by an arbitrary figure. \\
II. The lattice of Fig. 6 is bounded at $\mu_{ min}$ by an arbitrary
figure, and at $\mu_{max}$ by a segment parallel to the axis $j$. In any
other case the multiplicity will be greater than 1, equal to the
minimum number of factors $\alpha$ required to initiate the recurrence 
relationships. \\ \indent
As an example, in Fig. 6 is represented the diagram associated to the
highest weight in the case $P=P_1=P_2$, $Q= Q_1=Q_2$. It is easy to see
that the multiplicity of the representation $D(P,Q)$ in the decomposition
of the direct product is equal to the number of nodes of the lattice $A$
placed on the $j$ - axis. If $(P-Q)/3$ is an integer, this number
represents the multiplicity of the weight $i_3=0$, $y=0$ from the space
$V(P,Q)$, which is $1+ \min(P,Q)$. By this, one recovers the result
stated in ref. \cite{28}. \\ 
2. The highest weight vector $\vert s> \equiv \vert P~Q~
\underline{i} ~ \underline{i} ~ y > $ given by Eq. (194) - (197) depends
in general on free parameters which cannot be specified by the $SU(3)$ 
symmetry alone. In principle, it is possible to chose these
parameters such that to specify an orthonormal basis in $W(P,Q)$. However,      
the result has a physical meaning  only if there is a Hermitian operator
associated to an observable, commuting with all the operators of the 
set (b),  which has a diagonal matrix in this orthonormal basis. \\
3. Every point of the lattice of Fig. 3 is associated by Eq. (186) to a 
row in the matrix of the isoscalar factors. The calculation of the
factors $\alpha^s_{ \mu jk}$ along the row $( \underline{i},
\underline{y})$ was presented at step 1. The factors $\alpha^{iys}_{\mu j
k}$ from other rows can be related to $\alpha^s_{\mu jk}$ by an additional
set of recurrence formulas. \\ \indent 
Acting by $L_-$ on both sides of Eq. (186) it is possible to find
a recurrence relation between the factors placed on lines parallel
to $d$, 
%eq199
\begin{equation}
\sqrt{ \frac{ 2i+2}{i + j + k +2  } } \kappa^{ s}_{i +
\frac{1}{2} ~ y  -1}  \alpha^{  i+ \frac{1}{2} ~ y-1~s}_{ \mu  ~ j  ~ k +
\frac{1}{2} } = 
\sqrt{ \frac{ i+j-k}{ 2 j  } } \kappa^{ s_1}_{j  \mu }  
\alpha^{iys}_{ \mu +1 ~ j - \frac{1}{2} ~ k + \frac{1}{2} }
\end{equation}
$$
+ \sqrt{ \frac{ (i+k-j+1)(j+k-i+1)}{(2j+1)(i + j + k +2)  } } \chi^{
s_1}_{j  \mu }  \alpha^{ i  y s}_{ \mu +1  ~
j + \frac{1}{2}  ~ k +
\frac{1}{2} } 
$$
$$
+ \sqrt{ \frac{ i+k-j+1}{2k+1  } } \kappa^{ s_1}_{k +
\frac{1}{2} ~ y  - \mu -1}  \alpha^{ i  ys}_{ \mu j k }  
$$      
$$
- \sqrt{ \frac{ (i-k+j)(j+k-i+1)}{2(k+1)(i + j + k +2)  } } \chi^{
s_2}_{k + \frac{1}{2}  ~ y- \mu -1}  \alpha^{ i  ys }_{ \mu   ~
j   ~ k +1 }~~.
$$      
By using this formula it is possible to calculate successively the
isoscalar factors $\alpha^{iys}_{\mu jk}$ for any $(i,y)$, if the factors
along the lines parallel to $u$ or $\nu$ are known. \\ \indent
By acting with $K_-$ on both sides of Eq. (186) one arrives at
%eq200
\begin{equation}
\sqrt{ \frac{ (i+k -j+1)(j+k-i+1)(2i+2)}{(i + j + k +2)(2i+1)^2  } }
\kappa^{ s}_{i + \frac{1}{2} ~ y  -1}  \alpha^{ i+ \frac{1}{2} ~ y-1~s}_{
\mu  ~ j  ~ k + \frac{1}{2} } 
\end{equation}
$$ 
+ \sqrt{ \frac{ i+j-k}{ 2 i+1  } } \chi^{ s}_{i- \frac{1}{2} ~ y-1 }  
\alpha^{i-\frac{1}{2}~y-1~s}_{ \mu ~ j  ~ k + \frac{1}{2} }
$$
$$
= 
\sqrt{ \frac{ 2j+1}{i + j + k +2  } } \chi^{ s_1}_{j  \mu }  \alpha^{
i ys}_{ \mu +1  ~ j + \frac{1}{2} ~ k +
\frac{1}{2} } + 
\sqrt{ \frac{ j+k-i+1}{ 2 k+1  } } \kappa^{ s_2}_{k+ \frac{1}{2} ~ y- \mu
-1 } \alpha^{iys}_{ \mu   j   k  }
$$
$$
+ \sqrt{ \frac{ (i+j+k+1)(i+j-k)}{ 2( k+1)(i+j+k+2)  } } \chi^{
s_2}_{k+ \frac{1}{2} ~ y- \mu - 1 }  
\alpha^{iys}_{ \mu ~ j  ~ k + 1 }~~.
$$
By eliminating the factor $\alpha^{ i+ \frac{1}{2} ~ y-1~s}_{\mu ~ j ~ k +
\frac{1}{2}} $ between Eq. (199) and (200) one obtains a recurrence
relation on lines parallel to $u$, 
%eq201
\begin{equation}
 \chi^{ s}_{i- \frac{1}{2} ~ y - 1 }  
\alpha^{i- \frac{1}{2} ~y-1~s}_{ \mu ~ j  ~ k + \frac{1}{2} }=
- \sqrt{ \frac{ (i-j+k+1)(j+k-i+1)}{ 2j(2i+1)  } } \chi^{
s_1}_{j  \mu } \alpha^{iys}_{ \mu +1 ~ j- \frac{1}{2}  ~ k +
\frac{1}{2} }
\end{equation}
$$
+ \sqrt{ \frac{ (i+j+k+2)(i+j-k)}{ ( 2i+1)(2j+1)  } } \chi^{
s_1}_{j  \mu }  \alpha^{iys}_{ \mu +1 ~ j + \frac{1}{2}  ~ k +
\frac{1}{2} }
$$
$$
+ \sqrt{ \frac{ (i+j-k)(j+k-i+1)}{ ( 2i+1)(2k+1)  } } \kappa^{
s_2}_{k + \frac{1}{2} ~y-  \mu -1}  \alpha^{iys}_{ \mu  j   k }
$$
$$
+ \sqrt{ \frac{ (i+j+k+2)(i-j+k+1)}{ ( 2i+1)(2k+2)  } } \chi^{
s_2}_{k+ \frac{1}{2}  ~ y- \mu -1}  \alpha^{iys}_{ \mu  ~ j  ~ k +1 }~~.
$$
The equations (199) and (201) determine completely the isoscalar factors
$\alpha^{iys}_{ \mu jk}$ as a function of
$\alpha^{ \underline{i} ~ \underline{y}~s}_{ \mu jk} \equiv 
\alpha^s_{ \mu  j  k}$ (the indices $s_1$, $s_2$ of $\alpha$ appearing
in Eq. (186) have been omitted). 
The range of the indices $\mu$, $j$, $k$, 
as well as the shape of the lattice associated to each row in the axis
frame $(k, \mu, j)$, are given by Eq. (189) and (190).  In terms of
this lattice, Eq. (199) and (201) can be pictured geometrically by the 
structure represented in Fig. 8, as
expressing an unknown factor $\alpha$ at the point $P:(k+1/2, \mu,j)$ as
a function of known factors at the points 1,2,3,4. \\ \indent
In certain particular situations it is convenient to use also the
recurrence formula obtained by acting on both sides of Eq. (186) with the
operator $F$, constructed with the generators of Eq. (183).   This formula
is
%eq202
\begin{equation}
\rho^{s_1s_2}_{ \mu j  k} \alpha^{s iy}_{ \mu j k} = 
\lambda^{s_1s_2}_{ \mu jk} \alpha^{iys}_{\mu-1 ~ j- \frac{1}{2} ~
k- \frac{1}{2} } + 
\lambda^{s_1s_2}_{ \mu +1 ~j+ \frac{1}{2} ~k+ \frac{1}{2}}
\alpha^{iys}_{\mu +1 ~ j+ \frac{1}{2} ~
k+ \frac{1}{2} } 
\end{equation}
$$
- \tau^{s_1s_2}_{ \mu jk} \alpha^{iys}_{\mu-1 ~ j+ \frac{1}{2} ~
k+ \frac{1}{2} } -
\tau^{s_1s_2}_{ \mu +1 ~j- \frac{1}{2} ~k- \frac{1}{2}}
\alpha^{iys}_{\mu +1 ~ j- \frac{1}{2} ~
k- \frac{1}{2} }
$$
$$
+ \xi^{s_1s_2}_{ \mu jk} \alpha^{iys}_{\mu-1 ~ j- \frac{1}{2} ~
k+ \frac{1}{2} } +
\xi^{s_1s_2}_{ \mu +1 ~j+ \frac{1}{2} ~k- \frac{1}{2}}
\alpha^{iys}_{\mu +1 ~ j+ \frac{1}{2} ~
k- \frac{1}{2} }
$$
$$
+ \omega^{s_1s_2}_{ \mu jk} \alpha^{iys}_{\mu-1 ~ j+ \frac{1}{2} ~
k- \frac{1}{2} } + \omega^{s_1s_2}_{ \mu +1 ~j- \frac{1}{2}
~k+ \frac{1}{2}}
\alpha^{iys}_{\mu +1 ~ j- \frac{1}{2} ~
k+ \frac{1}{2} }
$$
where
$$
\rho^{s_1s_2}_{ \mu j k} = f_s - f_{s_1} - f_{s_2} - \frac{3}{2} (y-
\mu) - 2j(i-j) + (i+k-j+1)(j+k-i)
$$
$$
\lambda^{s_1s_2}_{ \mu jk} = \sqrt{ \frac{(i+j+k+1)(j+k-i)}{4jk}}
\chi^{s_1}_{j- \frac{1}{2} ~ \mu -1} \kappa^{s_2}_{k ~ y - \mu}
$$
$$
\tau^{s_1s_2}_{ \mu jk} = \sqrt{ \frac{(i+j+k+2)(j+k-i+1)}{(2j+1)(2k+1)}}
\kappa^{s_1}_{j+ \frac{1}{2} ~ \mu -1} \chi^{s_2}_{k ~ y - \mu}
$$
$$
\xi^{s_1s_2}_{ \mu jk} = \sqrt{ \frac{(i+j-k)(i+k-j+1)}{2j(2k+1)}}
\chi^{s_1}_{j- \frac{1}{2} ~ \mu -1} \chi^{s_2}_{k ~ y - \mu}
$$
$$
\omega^{s_1s_2}_{ \mu jk} = \sqrt{ \frac{(i-j+k)(i+j-k+1)}{2k(2j+1)}}
\kappa^{s_1}_{j+ \frac{1}{2} ~ \mu -1} \kappa^{s_2}_{k ~ y - \mu}
$$ 
and $f_s$ represents the eigenvalue of the operator $F$ corresponding to
the state with $s = (P,Q)$, given by Eq. (55). This recurrence formula
acts within the frame $( k, \mu , j)$ associated to a single line $(i,
y)$, and relates the isoscalar factors from the points represented in
Fig. 9.   \\[.5cm]
{\bf X. The canonical base of the space $V(P,Q)$ } \\[.5cm]   \indent
The canonical basis of the space $V(P,Q)$ was constructed in sect. VII
by direct product, and has the expression of Eq. (128). The unknown
isoscalar factors $\alpha^{iy (PQ)}_\mu \equiv \alpha^{
i~y~(PQ)~(P0)~(0Q)}_{\mu~ \frac{P}{3} + \frac{ \mu}{2} ~ \frac{Q}{3} - 
\frac{ y - \mu}{2}}$ appearing in this formula can be determined by using
Eq. (202). In this case $s_1 = (P,0)$, $s_2=(0,Q)$, and the 3D lattice of
Fig. 6 reduces to a one-dimensional lattice. Therefore, 
$\kappa^{s_1}=0$, $\chi^{s_2}=0$, and Eq. (202) takes the simple form
%eq203
\begin{equation}
\rho^{PQ}_\mu \alpha^{i~y~(PQ)}_\mu = 
\lambda^{PQ}_\mu \alpha^{i~y~(PQ)}_{\mu -1} + 
\lambda^{PQ}_{\mu +1} \alpha^{i~y~(PQ)}_{\mu +1} ~~,
\end{equation}
where
%eq204
\begin{equation} 
\rho^{PQ}_\mu = 
( \frac{P+Q}{3}+ i - \frac{y}{2} + \mu +1)( \frac{P+Q}{3}
-i - \frac{y}{2} + \mu)  
\end{equation}
$$ 
+ ( \frac{P}{3} - \mu )( \frac{Q}{3} + y - \mu ) ~~,
$$
%eq205
\begin{equation}
\lambda^{PQ}_\mu = \sqrt{ 
 \frac{P+Q}{3}+ i - \frac{y}{2} + \mu +1} 
\end{equation}
$$
\times \sqrt{( \frac{P+Q}{3} -i- \frac{y}{2} + \mu)( \frac{P}{3} - \mu + 1)(
\frac{Q}{3} + y - \mu +1) }  ~~.
$$  
By using the notations
%eq206
\begin{equation}
a_\mu = \frac{P+Q}{3} + i - \frac{y}{2} + \mu + 1 ~~,
\end{equation}
$$
b_\mu = \frac{P+Q}{3} - i - \frac{y}{2} + \mu ~~,
$$
and
$$
c_\mu = \frac{P}{3} - \mu ~~,~~ d_\mu = \frac{Q}{3} + y - \mu ~~,
$$
Eq. (203) takes the form
%eq207
\begin{equation}
a_\mu b_\mu \alpha_\mu^{iy(PQ)} + c_\mu d_\mu \alpha_\mu^{i y (PQ)} =
\sqrt{ a_\mu b_\mu c_{\mu-1} d_{\mu-1} } \alpha^{iy(PQ)}_{\mu-1} 
\end{equation}
$$
+
\sqrt{ a_{ \mu+1} b_{ \mu+1} c_\mu d_\mu } \alpha_{\mu+1}^{iy(PQ)} ~~.
$$
This equality is satisfyed if
%eq208
\begin{equation}
\sqrt{ a_\mu b_\mu } \alpha_\mu^{iy(PQ)} =
\sqrt{ c_{\mu-1} d_{\mu-1} } \alpha^{iy(PQ)}_{\mu-1} ~~,
\end{equation}
which is a recurrence relation with the solution
%eq209
\begin{equation}
\alpha^{iy (PQ)}_\mu = \alpha^{iy (PQ)}_{ \tilde{ \mu} }  
\end{equation}
$$
\times \sqrt{ \frac{ 
( \frac{P+Q}{3} + i - \frac{y}{2} +1 + \tilde{ \mu} ) !
( \frac{P+Q}{3} - i - \frac{y}{2} + \tilde{ \mu} ) !
( \frac{P}{3} - \tilde{ \mu} )! 
( \frac{Q}{3}+y - \tilde{ \mu} )! }{
( \frac{P+Q}{3} + i - \frac{y}{2} +1 +  \mu ) !
( \frac{P+Q}{3} - i - \frac{y}{2} +  \mu ) !
( \frac{P}{3} - \mu )! 
( \frac{Q}{3}+y -  \mu )! } } ~~.
$$
The maximum value $\tilde{ \mu}$ of $\mu$ is determined only by Eq. (189),
and is equal to $( \mu)_{max}$ given by Eq. (136). The minimum value of
$\mu$ is
$$
( \mu)_{min} = \max ( i + \frac{y}{2} - \frac{P+Q}{3} , y - 2 \frac{Q}{3}
, - 2 \frac{P}{3} ) ~~. 
$$
On the rows through the nodes of the line $d$ one obtains 
$( \mu)_{max} = ( \mu)_{min} = P/3$, and the sum over $\mu$ in Eq. (128)
reduces to a single term,
%eq210
\begin{equation}
\vert P~ Q ~ \Delta ~ i_3 ~ \tau> = \alpha^{ \Delta ~ \tau ~ (PQ)}_{
\frac{P}{3} } \sum_{m} C^{ \frac{P}{2} ~ \Delta - \frac{P}{2} ~
\Delta }_{ m~ i_3-m~ i_3}~ \xi^{ \frac{P}{3}}_{Pm} \eta^{ \tau -
\frac{P}{3}}_{Q ~i_3-m} ~~.
\end{equation}
The unitarity condition implies $\vert \alpha^{ \Delta  \tau  (PQ)}_{
\frac{P}{3}} \vert =1$, and the action of $L_+$ on the vector 
$\vert P~Q~ \Delta ~ \Delta ~ \tau >$ shows  that all these factors have
the same sign, and can be considered equal to 1. The factors 
$\alpha^{ iy (PQ)}_{ \frac{P}{3}} $ can be calculated by using
Eq. (201), which takes the form
%eq211
\begin{equation}
 \chi^{ PQ}_{i- \frac{1}{2} ~ y - 1 }  
\alpha^{i- \frac{1}{2} ~y-1~(PQ)}_{ \mu  }=
 \sqrt{ \frac{ (i+j+k+2)(i+j-k)}{ ( 2i+1)(2j+1)  } } \chi^{
P}_{j  \mu }  \alpha^{iy(PQ)}_{ \mu +1 }
\end{equation}
$$
+ \sqrt{ \frac{ (i+j-k)(j+k+i-1)}{ ( 2i+1)(2k+1)  } } \kappa^{
Q}_{k + \frac{1}{2} ~y-  \mu -1}  \alpha^{iy(PQ)}_{ \mu }~~.
$$
In the case $y \ge \frac{P-Q}{3}$, for $\mu = P/3$ this equality becomes
%eq212
\begin{equation}
\sqrt{ \frac{P+2Q}{3} + i + \frac{y}{2} + 1 } ~ \alpha^{i - \frac{1}{2} ~
y-1 ~ (PQ)}_{ \frac{P}{3}} = \sqrt{ y - \frac{P-Q}{3}} ~ \alpha^{
i~y~(PQ)}_{\frac{P}{3}} ~~.
\end{equation}
Finally, by solving this recurrence relation with the initial condition
$\alpha^{ \Delta \tau (PQ)}_{P/3} = 1$, one obtains  
%eq213
\begin{equation}
\alpha^{ iy  (PQ)}_{ \frac{P}{3}} = 
\sqrt{ \frac{ 
( \frac{P+2Q}{3} + i + \frac{y}{2} +1 )!
( \frac{P+2Q}{3} - i + \frac{y}{2}  )!}{
(P+Q+1)! ( \frac{Q-P}{3} +y)!
} } ~~.
\end{equation}
In the case $y \le \frac{P-Q}{3}$, by using Eq. (208) and (211) one can
find similarly  
%eq214
\begin{equation}
\alpha^{ iy  (PQ)}_{y+  \frac{Q}{3}} = 
\sqrt{ \frac{ 
( \frac{2P+Q}{3} - i - \frac{y}{2} )!
( \frac{2P+Q}{3} + i - \frac{y}{2} +1 )!}{
(P+Q+1)! ( \frac{P-Q}{3} - y)!
} } ~~,
\end{equation}
such that in both cases, Eq. (209) takes the form
%eq215 
\begin{equation}
\alpha^{iy (PQ)}_\mu =   \sqrt{ \frac{ 
( \frac{P+2Q}{3} + i + \frac{y}{2} +1 ) !}{(P+Q+1)!} } 
\end{equation}
$$
\times \sqrt{ \frac{ ( \frac{P+2Q}{3} - i + \frac{y}{2}  ) !
( \frac{2P+Q}{3} - i - \frac{y}{2} )! 
( \frac{2P+ Q}{3}+ i - \frac{y}{2} + 1 )! }{
( \frac{P+Q}{3} + i - \frac{y}{2} +1 +  \mu ) !
( \frac{P+Q}{3} - i - \frac{y}{2} +  \mu ) !
( \frac{P}{3} - \mu )! 
( \frac{Q}{3}+y -  \mu )! } } ~~.
$$
One can check that the factors determined by this equation 
have the important symmetry property
%eq216
\begin{equation}
\alpha^{ iy(PQ)}_\mu = \alpha^{ i ~ -y ~ (QP)}_{ \mu -y}~~.
\end{equation}
This property will be used in the following to prove the equivalence
between the representations $D(P,Q)^*$ and $D(Q,P)$. By using Eq. (128)
one obtains 
%eq217
\begin{equation}
\vert QP~ i~ -i_3 ~-y> = \sum_{ \mu m } 
\alpha^{ i~- y~ (QP) }_\mu   
C^{ j  ~ k ~ i}_{m~ -i_3-m ~- i_3} \xi^\mu_{Qm} \eta^{ - y - \mu}_{P
- i_3-m} ~~.
\end{equation}
If the basis elements $\xi^\mu_{Q~m}$ and $\eta^{ - y - \mu}_{ P ~ -
i_3-m}$ appearing here are expressed in terms of $ ( \eta^{ - \mu}_{Q ~
-m})^*$, respectively $( \xi^{y+ \mu}_{P~ i_3 +m} )^*$, by using
Eq. (100),  and then we introduce the new summation indices
$n = i_3 +m$ and $\nu = y+ \mu$,  then Eq. (217) becomes
%eq218
\begin{equation}
\vert QP~ i~ - i_3 ~-y> = \sum_{ \mu n } 
\alpha^{ i~- y~ (QP) }_{\nu- y}    
\end{equation}
$$
\times C^{ j  ~ k ~ i}_{n- i_3~ -n ~- i_3} 
(-1)^{ \frac{P-Q}{3} + i_3 + \frac{y}{2}}
( \xi^\nu_{Pn} \eta^{ y - \nu}_{Q
i_3-n})^* ~~.
$$
By using now Eq. (216), and the symmetry properties of the $SU(2)$
Clebsch-Gordan coefficients,
%eq219
\begin{equation}
C^{  ~ j ~~ k ~~ i}_{ n-i_3 ~ -n ~ -i_3} = C^{ k ~ j ~ i}_{n~i_3-n~i_3}~~,
\end{equation}
one arrives at
%eq220
\begin{equation}
 \vert QP~i ~ -i_3~ -y> = (-1)^{ \frac{P-Q}{3} + i_3 + \frac{y}{2}} 
\vert PQ~ i~i_3~y>^* ~~. 
\end{equation}
The sign of this expression is given by the number $t$, called
triality \cite{29}, defined as
%eq221
\begin{equation}
t = (P-Q)~ mod~3~~,~~ t = -1,0,1
\end{equation}
such that up to a negligible global phase factor, Eq. (220)  becomes
%eq222
\begin{equation}
 \vert QP~i ~ -i_3~ -y> = (-1)^{ \frac{t}{3} + i_3 + \frac{y}{2}} 
\vert PQ~ i~i_3~y>^* ~~. 
\end{equation}
\indent The $SU(3)$ representations which map the elements of the
discrete subgroup $Z_3$ onto the unity matrix $I$ are faithful
representations of the factor group $SU(3)/Z_3$. For these representations
the integers $m_1$, $m_2$, and $m_3$ of Eq. (42) should satisfy the
equality
%eq223
\begin{equation}
(m_1+m_2+m_3)~ mod~ 3 = (P-Q)~ mod~3=0 ~~,
\end{equation}
which means that both the electric charge $i_3+ y/2$ which appears in
Eq. (222), and the hypercharge $y$, are integers. \\[.5cm]
{\bf XI. The symmetric form of the canonical basis in the spaces
$V(P,0)$ and $V(0,Q)$. } \\[.5cm]
\indent The vectors $\xi^y_{Pi_3}$ of the canonical basis of the space
$V(P,0)$  can be expressed as a symmetrized product of the 
basis elements of $V(3,0)$, according to Eq. (98),
%eq224
\begin{equation}
\xi^y_{P i_3}  = 
a^P(i_3,y)  (x_1)^{i+i_3}(x_2)^{i-i_3}(x_3)^{ \frac{P}{3}-y}~~,~~
i= \frac{P}{3} + \frac{y}{2}~~.
\end{equation} 
The unknown factors $a^P(i_3,y)$ can be calculated by recurrence.
Acting on both sides of Eq. (224) with the operator $I_+$ one obtains
%eq225
\begin{equation}
a^P(i_3,y) = \sqrt{ \frac{i+i_3+1}{i-i_3}} a^P(i_3+1,y)~~,
\end{equation}
which has the solution
%eq226
\begin{equation}
a^P(i_3,y) = \sqrt{ \frac{ (2i)!}{(i-i_3)!(i+i_3)!}} a^P_y~~,
\end{equation}
where $a^P_y \equiv a^P(i,y)$. By choosing in Eq. (224) $i_3=i$,
and acting on both sides with $K_+$, one ontains
%eq227
\begin{equation}
a^P_y= \sqrt{ \frac{ \frac{P}{3} -y+1}{2 \frac{P}{3} + y}} a^P_{y-1}~~.
\end{equation}
By recurrence, this gives
%eq228
\begin{equation}
a^P_y = \sqrt{ \frac{P!}{ ( \frac{2P}{3} + y )!( \frac{P}{3} -y)! } }
~a^P_{ \frac{P}{3}} =
\end{equation}
$$ 
\sqrt{ \frac{P!}{ ( 2i )!( \frac{P}{3} -y)! } }
~a^P_{ \frac{P}{3}} ~~.
$$
Then, by using Eq. (226) with $a^P_{P/3}=1$, one obtains
%eq229
\begin{equation}
a^P(i_3,y) =  \sqrt{ \frac{P!}{ (i-i_3)!(i+i_3)!
( \frac{P}{3} -y)! } } ~~.
\end{equation}
The symmetric basis of the space $V(0,Q)$ is expressed by Eq. (103),
where all the quantities are now determined. \\[.5cm]
{\bf XII. The isoscalar factors} 
$\alpha^{ iy (P_1+P_2,0)(P_1,0)(P_2,0)}_{\mu~j~k}$ 
{\bf and } 
$\alpha^{ iy (0,Q_1+Q_2)(0,Q_1)(0,Q_2)}_{\mu~j~k}$  \\[.5cm]
\indent If $s_1= (P_1,0)$, $s_2=(P_2,0)$,  then $\kappa^{s_1} =
\kappa^{s_2}=0$, and Eq. (202) takes the simple form
%eq230
\begin{equation}
\rho^{P_1P_2}_{ \mu} \alpha^{y(P_1+P_2,0)}_{ \mu} = 
\xi^{P_1P_2}_{ \mu } \alpha^{y(P_1+P_2,0)}_{\mu-1 } + 
\xi^{P_1P_2}_{ \mu +1}
\alpha^{y(P_1+P_2,0)}_{\mu}~~, 
\end{equation}
where $\alpha^{ y (P_1+P_2,0)}_\mu \equiv \alpha^{
iy(P_1+P_2,0)~(P_1,0)~(P_2,0)}_{ \mu ~j~k}$, 
$$
j= \frac{P_1}{3} + \frac{ \mu}{2}~~,~~ k = \frac{P_2}{3} + \frac{y-
\mu}{2} ~~,~~ i = \frac{P_1+P_2}{3} + \frac{y}{2} ~~,
$$
$$
\rho^{P_1P_2}_\mu = ( \frac{2P_1}{3}+ \mu)( \frac{P_2}{3} - y + \mu) + (
\frac{P_1}{3} - \mu)( \frac{2P_2}{3} + y - \mu)~~,
$$
and
$$
\xi^{P_1P_2}_\mu = \sqrt{
( \frac{P_1}{3}- \mu +1)( \frac{P_2}{3} - y + \mu) (
\frac{2P_1}{3} + \mu)( \frac{2P_2}{3} + y - \mu+1) } ~~.
$$
The extreme values of the index $\mu$ are given by Eq. (189), as
%eq231
\begin{equation}
\max ( - 2 \frac{P_1}{3} , y - \frac{P_2}{3} ) \le \mu \le
\min( \frac{P_1}{3}, y + 2 \frac{P_2}{3} )~~,
\end{equation}
because in this case $i=j+k$.  \\ \indent
Following the same steps as in sect. XI, it is possible to show that the
solutions of Eq. (230) satisfy the relationship
%eq232
\begin{equation}
\sqrt{ ( \frac{P_1}{3} - \mu)( \frac{2P_2}{3} + y - \mu) } 
\alpha_\mu^{ y ~ (P_1+P_2,0)} = 
\end{equation}
$$
= \sqrt{ ( \frac{2P_1}{3} + \mu+1)( \frac{P_2}{3} - y + \mu +1) } 
\alpha_{\mu+1}^{ y ~ (P_1+P_2,0)} ~~,
$$
which gives
%eq233
\begin{equation}
\alpha^{ y ~(P_1+P_2,0)}_\mu = \alpha^{ y (P_1+P_2,0)}_{ \tilde{
\mu}} \times
\end{equation}
$$
\sqrt{ \frac{ 
( \frac{P_1}{3} - \tilde{ \mu} )! 
( \frac{2P_2}{3} +y- \tilde{ \mu} )! 
( \frac{2P_1}{3} + \tilde{ \mu} )! 
( \frac{P_2}{3} -y+ \tilde{ \mu} )! }{
( \frac{P_1}{3} -  \mu )! 
( \frac{2P_2}{3} +y-  \mu )! 
( \frac{2P_1}{3} + \mu )! 
( \frac{P_2}{3} -y+  \mu )! }
} ~~.
$$
Here $\tilde{ \mu} = ( \mu)_{max} = P_1/3$ if $y \ge (P_1-2P_2)/3$ and
$\tilde{ \mu} = y + 2P_2/3$ if $y \le (P_1-2P_2)/3$. To calculate
$\alpha^{ y~ (P_1+P_2,0)}_\mu$ one can use Eq. (201), which in this
case takes the form
%eq234
\begin{equation}
\sqrt{ ( \frac{P_1+P_2}{3} - y+1)(2 \frac{P_1+P_2}{3} + y ) } 
\alpha_\mu^{ y-1 ~ (P_1+P_2,0)}  
\end{equation}
$$
= \sqrt{ ( \frac{2P_1}{3} + \mu+1)( \frac{P_1}{3} - \mu) } 
\alpha_{\mu+1}^{ y ~ (P_1+P_2,0)}
$$ 
$$
+ \sqrt{ ( \frac{2P_2}{3} +y- \mu)( \frac{P_2}{3} -y + \mu +1) } 
\alpha_{\mu}^{ y ~ (P_1+P_2,0)}~~.
$$ 
By using the initial condition $\alpha^{ \frac{P_1+P_2}{3}~(P_1+P_2,0)}_{ 
\frac{P_1}{3}}=1$, one obtains
%eq235
\begin{equation}
\alpha_{ \frac{P_1}{3}}^{ y ~ (P_1+P_2,0)}  
= \sqrt{ \frac{P_2! (2 \frac{P_1+P_2}{3} + y)! }{ 
(P_1+P_2)! ( \frac{2P_2-P_1}{3}+ y)!} }
\end{equation}
if
$$
y \ge \frac{P_1-2P_2}{3}
$$
and
%eq236
\begin{equation}
\alpha_{ \frac{2P_2}{3} +y}^{ y ~ (P_1+P_2,0)}  
= \sqrt{ \frac{P_1! ( \frac{P_1+P_2}{3} - y)! }{ 
(P_1+P_2)! ( \frac{P_1-2P_2}{3}- y)!} }
\end{equation}
if
$$
y \le \frac{P_1-2P_2}{3}~~.
$$
In both cases, the isoscalar factors given by Eq. (233) are
%eq237
\begin{equation}
\alpha^{ \mu ~(P_1+P_2,0)}_\mu = 
\end{equation}
$$
\sqrt{ \frac{ P_1! P_2! 
( \frac{P_1+P_2}{3} - y )! 
(2 \frac{P_1+P_2}{3} +y )! }{ (P_1+P_2)!
( \frac{P_1}{3} -  \mu )! 
( \frac{2P_2}{3} +y-  \mu )! 
( \frac{2P_1}{3} + \mu )! 
( \frac{P_2}{3} -y+  \mu )! }
} ~~.
$$
By taking the complex conjugate of the equality 
%eq238
\begin{equation}
\xi^y_{Q_1+Q_2~i} = \sum_\mu \alpha^{y (Q_1+Q_2,0)}_\mu \xi^\mu_{Q_1 j}
\xi^{y-\mu}_{Q_2~i-j}~~,
\end{equation}
and making use of Eq. (100), one obtains
%eq239
\begin{equation}
\eta^y_{Q_1+Q_2~-i} = \sum_\mu \alpha^{-y~(Q_1+Q_2,0)}_{- \mu} 
\eta^\mu_{Q_1~-j} \eta^{y- \mu}_{Q_2~-(i-j)}~~.
\end{equation}
This result allows to express the isoscalar factors $\alpha^{ y (0,
Q_1+Q_2)}_\mu$ in terms of the ones given by Eq. (237), by identifying
summation terms in Eq. (239) and (186) for the case
$s_1=(0,Q_1)$, $s_2=(0,Q_2)$, $s=(0,Q_1+Q_2)$, $i_3=i=j+k$, such that
%eq240
\begin{equation}
\alpha^{ y~(0,Q_1+Q_2)}_\mu = \alpha^{ -y~(Q_1+Q_2,0)}_{- \mu} ~~.
\end{equation}
{\bf XIII. The isoscalar factors } $\alpha^{00~(0,0)(P,Q)(Q,P)}_{ \mu~j~k}
$ \\[.5cm]
\indent These isoscalar factors can be obtained by using the 
conjugation relationship of Eq. (222) in the expression of 
the normalized $SU(3)$ scalar which can be constructed 
with the elements of the canonical base of the space $V(P,Q)$. 
This scalar is
%eq241
\begin{equation}
S= \frac{1}{\sqrt{ \dim V(P,Q) }} \sum_{ y i_3 j  k} \delta_{jk}
< PQ ~j~i_3~ y \vert PQ~k~i_3~y> = 
\end{equation}
$$
 \frac{1}{\sqrt{ \dim V(P,Q) }} \sum_{ y i_3 j  k} \delta_{jk}
(-1)^{- (\frac{t}{3} +m+ \frac{ \mu}{2}) }  \vert PQ ~j~ m~ \mu>  \vert
QP~k~-m~- \mu> ~~. 
$$
The $SU(2)$ Clebsch-Gordan coefficient in this case is
%eq242
\begin{equation}
C^{ j~~k~~0}_{m ~-m~0} = \delta_{j~k} \frac{ (-1)^{j-m} }{ \sqrt{2j+1}}~~,
\end{equation}
and therefore the isoscalar factor is
%eq243
\begin{equation}
\alpha^{00~(0,0)(P,Q)(Q,P)}_{\mu~j~j} = \sqrt{ \frac{ 2j+1}{ \dim V(P,Q)}}
~(-1)^{ - ( \frac{t}{3}+\frac{\mu}{2} +j)}~~.
\end{equation}
The maximum value of $j$ is $(P+Q)/2$, and for this value the sign of the
factor determined by  Eq. (243) is $(-1)^{ Q+t}$. When $Q+t$ is
odd, the factor is negative, violating the phase convention used in
ref. \cite{9}. However, all these factors are defined up to a constant, 
which can be chosen such that those with a maximum $j$
are positive. The general expression in this case is  
%eq244
\begin{equation}
\alpha^{00~(0,0)(P,Q)(Q,P)}_{\mu~j~j} = \sqrt{ \frac{2( 2j+1)}{
(P+1)(Q+1)(P+Q+2)}}
~(-1)^{ - ( \frac{2t}{3}+\frac{\mu}{2} +j+Q)}~~.
\end{equation} 
{\bf XIV. The highest weight vectors in the space} $V(1,1) \otimes
V(1,1)$. \\[.5cm]
\indent The Clebsch-Gordan series associated to this product has the form
%eq245
\begin{equation}
V(1,1) \otimes V(1,1) = V(2,2) + V(3,0)+V(0,3) +
\end{equation}
$$
+ V^1(1,1)+V^2(1,1) + V(0,0)~~.
$$
For the highest weight $( \underline{i} , \underline{y} ) =
(P/2, (P+2Q)/3 )$ of each space $V(P,Q)$ appearing in this sum we
can construct a diagram $( k, \mu, j)$ similar to the one pictured in 
Fig. 6. This gives a geometrical representation of the range of the
indices resulted from Eq. (196) and (197), making easier to apply
Eq. (194), (195) to each term of the sum. \\
(A). $V(2,2)$; $\underline{i} =1$, $\underline{y} =2$, $\mu=1$. 
There is one non-vanishing factor, 
$\alpha^{12(2,2)}_{1~\frac{1}{2}~\frac{1}{2}}$, and the highest weight 
vector is
%eq245
\begin{equation}
\vert 22~112> = 
\vert 11~\frac{1}{2} ~ \frac{1}{2} 1>   
\vert 11~\frac{1}{2} ~ \frac{1}{2} 1>~~.
\end{equation}
(B). $V(3,0)$; $\underline{i} =3/2$, $\underline{y} =1$, $\mu=0,1$. 
There are two non-vanishing factors,
$\alpha^{\frac{3}{2}~1~(3,0)}_{0~1~\frac{1}{2}}$, 
$\alpha^{\frac{3}{2}~1~(3,0)}_{1~ \frac{1}{2}~ 1}$,
and in this case  Eq. (195) becomes 
%eq246
\begin{equation}
\alpha^{\frac{3}{2}~1~(3,0)}_{0~1~\frac{1}{2}} +
\alpha^{\frac{3}{2}~1~(3,0)}_{1~ \frac{1}{2}~ 1} =0 ~~.
\end{equation}
This equation, combined with the unitarity condition
%eq247
\begin{equation}
(\alpha^{\frac{3}{2}~1~(3,0)}_{0~1~\frac{1}{2}})^2 +
(\alpha^{\frac{3}{2}~1~(3,0)}_{1~ \frac{1}{2}~ 1})^2 =1 ~~,
\end{equation}
and the phase convention, determines the values
%eq248
\begin{equation}
\alpha^{\frac{3}{2}~1~(3,0)}_{0~1~\frac{1}{2}}= \frac{1}{ \sqrt{2}} ~~,~~
\alpha^{\frac{3}{2}~1~(3,0)}_{1~ \frac{1}{2}~ 1} = - \frac{1}{ \sqrt{2}} 
\end{equation}
of the isoscalar factors, and the highest weight vector 
%eq250
\begin{equation}
\vert 30~ \frac{3}{2} ~ \frac{3}{2} ~1> = \frac{1}{ \sqrt{2}} 
\sum_m \lbrack 
C^{ 1~ \frac{1}{2} ~ \frac{3}{2} }_{m~ \frac{3}{2} -m~
\frac{3}{2}} \vert 1~1~1~m~0> \vert 1~1~\frac{1}{2}~\frac{3}{2} -m~1> 
\end{equation}
$$
-
C^{ \frac{1}{2} ~ 1  ~ \frac{3}{2} }_{m~ \frac{3}{2} -m~
\frac{3}{2}} \vert 1~1~ \frac{1}{2}~m~1> \vert 1~1~1~\frac{3}{2}
-m~0>  \rbrack ~~.
$$
(C). $V(0,3)$; $\underline{i} =0$, $\underline{y} =2$, $\mu=1$. 
There is one non-vanishing factor, 
$\alpha^{02(0,3)}_{1~\frac{1}{2}~\frac{1}{2}}$, of modulus $1$, and the
highest weight vector is
%eq251
\begin{equation}
\vert 03~002> = 
\alpha^{02~(0,3)}_{1~\frac{1}{2}~\frac{1}{2}} \sum_m
C^{ \frac{1}{2} ~ \frac{1}{2} ~ 0 }_{ m~-m~0}
\vert 11~\frac{1}{2} ~ m~  1>   
\vert 11~\frac{1}{2} ~ -m ~ 1>~~.
\end{equation}
The phase of this factor can be determined by noticing that from Eq. (250) 
we get
%eq252
\begin{equation}
\vert 30~ 0~0 ~ -2 > =  \sum_m  C^{ \frac{1}{2} ~ \frac{1}{2} ~ 0 }_{m~
-m~ 0} \vert 1~1~ \frac{1}{2}  ~m~ -1> \vert 1~1~\frac{1}{2}~ -m~-1> ~~,
\end{equation}
while from Eq. (222)
%eq252
\begin{equation}
 \vert 03~ 0~0~2>^*= - \vert 30~0~0~ -2>~~,
\end{equation}
such that $\alpha^{02~(0,3)}_{1~\frac{1}{2}~\frac{1}{2}}=-1$. \\
(D). $V^1(1,1)+ V^2(1,1)$; $\underline{i} = 1/2$, $\underline{y} =1$,
$\mu=0,1$. The general expression of the highest weight vector in this
case is 
%eq253
\begin{equation}
\vert 11~ \frac{1}{2} ~ \frac{1}{2} ~1> = \sum_{ \mu jk} 
\alpha^{ \frac{1}{2} ~1~(11)}_{ \mu j k} \times
\end{equation}
$$  
\sum_m  C^{ j~ k ~ \frac{1}{2} }_{m~ \frac{1}{2} -m~
\frac{1}{2}} \vert 11~j~m~ \mu> \vert 11~k~\frac{1}{2} -m~1- \mu>~~, 
$$
where it is necessary to determine the factors
$$
\alpha^{ \frac{1}{2} ~1~(11)}_{ 0 1  \frac{1}{2}}~~,~~  
\alpha^{ \frac{1}{2} ~1~(11)}_{ 0 0  \frac{1}{2}}~~,~~
\alpha^{ \frac{1}{2} ~1~(11)}_{ 1  \frac{1}{2}1} ~~,~~
\alpha^{ \frac{1}{2} ~1~(11)}_{ 1  \frac{1}{2}0}~~. 
$$  
The recurrence equations (194), (195) become in this case
%eq254
\begin{equation}
\alpha^{ \frac{1}{2} ~1~(11)}_{ 0 0  \frac{1}{2}} = - \frac{1}{2} (
\alpha^{ \frac{1}{2} ~1~(11)}_{ 1 \frac{1}{2} 0 } +
\alpha^{ \frac{1}{2} ~1~(11)}_{ 1  \frac{1}{2} 1} )~~,~~
\end{equation}
%eq255
\begin{equation}
\alpha^{ \frac{1}{2} ~1~(11)}_{ 0 1  \frac{1}{2}} =  \frac{1}{2} (
3 \alpha^{ \frac{1}{2} ~1~(11)}_{ 1 \frac{1}{2} 0 } -
\alpha^{ \frac{1}{2} ~1~(11)}_{ 1  \frac{1}{2} 1} )~~,
\end{equation}
which should be solved together with the unitarity condition
%eq257
\begin{equation}
\sum_{ \mu j k} \vert \alpha^{ \frac{1}{2} 1~ (11)}_{ \mu j k} \vert^2=1
~~.
\end{equation}
However, by contrast to the previous situations, this system of equations
does not has an unique solution. \\ \indent
A solution of the system is provided by the set used in ref. \cite{9}, 
%eq258
\begin{equation}
- \alpha^{ \frac{1}{2} ~1~(11)}_{ 0 0  \frac{1}{2}} =  
\alpha^{ \frac{1}{2} ~1~(11)}_{ 0 1  \frac{1}{2}} =
\alpha^{ \frac{1}{2} ~1~(11)}_{ 1  \frac{1}{2} 0} =
\alpha^{ \frac{1}{2} ~1~(11)}_{ 1  \frac{1}{2} 1} = \frac{1}{2}~~.
\end{equation}
An independent set, $\{ \beta^{\frac{1}{2} 1 (11)}_{ \mu j k} \}$, can be
determined by adding to the system of Eq. (255)-(257), the orthogonality
condition
%eq259
\begin{equation}
\sum_{ \mu j k}  \alpha^{ \frac{1}{2} 1~ (11)}_{ \mu j k} 
\beta^{ \frac{1}{2} 1~(11)}_{ \mu j k}
=0 ~~.
\end{equation}   
The solution of the system obtained is
%eq260
\begin{equation}
\beta^{ \frac{1}{2} ~1~(11)}_{ 0 0  \frac{1}{2}} =  
\alpha^{ \frac{1}{2} ~1~(11)}_{ 1  \frac{1}{2} 0} = - \frac{ \sqrt{5}}{10}
~~,~~
- \beta^{ \frac{1}{2} ~1~(11)}_{ 0 1  \frac{1}{2} } =
\beta^{ \frac{1}{2} ~1~(11)}_{ 1  \frac{1}{2} 1} = 3 \frac{
\sqrt{5} }{10}~~.
\end{equation}  
The factors $\alpha$ and $\beta$ given by Eq. (258) and (260) determine by
Eq. (254) the highest weight vectors for the irrep spaces $V^1(1,1)$ and
$V^2(1,1)$, respectively. \\ 
(E). $V(0,0)$; $\underline{i} = 0$, $\underline{y} =0$,
$\mu=-1,0,1$. This space is one-dimensional, and the  highest weight
vector is the normalized $SU(3)$ scalar $S$ given by Eq. (241) for
$P=Q=1$. Therefore, the isoscalar factors can be calculated by using 
Eq. (244), and have the values
%eq261
\begin{equation}
\alpha^{00(00)}_{ 011} = \frac{ \sqrt{6}}{4}~~,~~
\alpha^{00(00)}_{ -1 \frac{1}{2}  \frac{1}{2}} =- \frac{ 1}{2}~~,~~
\alpha^{00(00)}_{ 000} = - \frac{ \sqrt{2}}{4}~~,~~
\alpha^{00(00)}_{ 1 \frac{1}{2}  \frac{1}{2}} = \frac{ 1}{2}~~.
\end{equation}
The same result can be obtained by using the recurrence
formulas. \\[.5cm]
{\bf Concluding remarks} \\[.5cm]
\indent The expressions of the isoscalar factors given in
Eq. (215) and (237), derived here by using Eq. (202), are the same as
in ref. \cite{22}. Also, the isoscalar factors determined in sect. XIV
have the  values  given in the tables \cite{9}. \\ \indent
The symmetry properties of the $SU(3)$ Clebsch-Gordan coefficients 
can be found by using the symmetry properties of  the
recurrence relationships for the isoscalar factors. However, it is easier
to derive them from the integral representation of the Clebsch-Gordan
coefficients, given by the projection operators associated to the
matrix elements of the irreducible representations \cite{9}, \cite{28},
\cite{34}, \cite{35}. \\ \indent
In general, for the group $SU(n)$,  $n>2$, the canonical factorization
$SU(n) \supset U(1) \times SU(n-1)$ can only solve the problem 
of the multiplicity for the weights, but not for the representations
appearing in the Clebsch-Gordan series.  Therefore, whatever method is
used for reduction, it is not possible to obtain numerical values for all
the Clebsch-Gordan coefficients, without making additional assumptions.   
Biedenharn and Louck \cite{36} have shown that by using the general 
property of canonical embedding for $U(n)$,
%eq261
\begin{equation}
U(n^2) \supset U(n) \times U(n) \supset U(n)
\end{equation}
it is possible to solve the problem of the multiplicities such that all
Clebsch-Gordan coefficients are determined.  \\ \indent
The recurrence equations for the $SU(n)$ isoscalar factors obtained by
using the canonical factorization can be derived by acting on the weight 
vectors with the representation operators for the $sl(n,C)$ generators
associated  to the $n-1$ simple roots. The action of the $sl(n,C)$
operators associated to the remaining positive roots leads to an
additional set of
%eq262
\begin{equation}
\frac{n(n-1)}{2} - (n-1) = \frac{ (n-1)(n-2)}{2}
\end{equation}
equations, which are identically satisfied due to the properties of
the $SU(n-1)$ Clebsch-Gordan coefficients. Therefore, these
relationships may provide in general only the dependence of the isoscalar
factors on arbitrary  parameters, because for $n>3$, $SU(n-1)$ is not
simply reducible. In the treatment of a specific  quantum system, the use
of a complete set of compatible observables might eliminate the
ambiguities, leading to well-defined values of these parameters in each
physical situation.
 \\[.5cm] 
{\bf Acknowledgements.} Thanks are due to Professor Gerry McKeon, for
interest and the support provided towards the translation in
English of this article.   

\end{document}